\documentclass[12pt]{article}

\usepackage[margin=1in]{geometry}
\usepackage{amsfonts, amsmath, amssymb}
\usepackage[none]{hyphenat}
\usepackage{fancyhdr}
\usepackage{float}
\usepackage{framed} 
\usepackage[]{mdframed}

\usepackage{array}
\usepackage{lscape} 

\usepackage{listings} 
\usepackage{longtable}

\usepackage{url}
\usepackage{rotating}

\usepackage{natbib}

\begin{document}

\providecommand{\normaldistn}{\mathrm{Normal}}  
\providecommand{\betadistn}{\mathrm{Beta}}      
\providecommand{\binomdistn}{\mathrm{Bin}}      
\providecommand{\halfnormaldistn}{\mathrm{halfNormal}}  
\providecommand{\expect}{\mathrm{E}}            
\providecommand{\var}{\mathrm{Var}}             

\begin{titlepage}
\begin{center}

\vspace*{1cm}
\Large{\textbf{}}\\
\vfill
\line(1,0){450}\\
\Large{\textbf{Including historical control data in simultaneous inference for pre-clinical multi-arm studies}}\\

\line(1,0){450}\\
\vfill

\flushleft
\small
Max Menssen$^1$, Carsten Kneuer$^2$, Gyamfi Akyianu$^2$, Christian Röver$^1$, Tim Friede$^1$, Frank Schaarschmidt$^3$\\

1: Department of Medical Statistics, University Medical Center Göttingen, Göttingen, Germany\\
2: German Federal Institute for Risk Assessment, Department of Pesticide Safety, 
Berlin, Germany \\
3: Leibniz University Hannover, Department of Biostatistics, Germany

\end{center}
\end{titlepage}

\setcounter{page}{1}


\section*{Keywords}

Bayesian borrowing, EFSA scientific opinion, virtual control group, 
long-term carcinogenicity study, pre-clinical risk assessment,
meta-analytic predictive prior distribution


\section*{Abstract}

In pre- and non-clinical toxicology, the reduction of animal use is highly desireable. Although
approaches for possible sample size reduction in the concurrent control group
were suggested previously under the virtual control groups framework for continuous endpoints, 
methodology that is applicable to binary outcomes that occur in long-term carcinogenicity
studies is currently missing. 
In order to augment animals in the current control group with historical control data,
we propose approaches that rely on dynamic Bayesian borrowing and simultaneous credible
intervals for risk ratios. Several operation characteristics such as familywise
error rate (FWER) and power are assessed via Monte-Carlo simulations and compared to 
the ones of approaches that rely on pooling of historical and 
current observations. 
It turned out that under optimal conditions, Bayesian approaches based on robustified prior distributions enable a substantial reduction of the control groups sample size, while still controlling the FWER up to a satisfactory level.
Furthermore, at least to some extend, 
these approaches were able to protect against possible drift. This hightlights the
potential of Bayesian study designs to reduce animal use in toxicology through 
re-use of the large pool of existing control data.

\newpage


\section{Introduction}
According to the 3R principle (Replace, Reduce, Refine), the reduction of animal use in toxicological studies 
is a key objective. For in vivo tests that remain unavoidable at the time being, 
this may be achieved through novel trial designs in which the sample 
size of the concurrent control group (CCG) is reduced, 
while the spared animals are augmented using external information, such as historical 
control data (HCD) (Steger-Hartmann et al. 2020). 

Indeed, it seems that the potential for animal reduction in pre-clinical toxicology is not 
fully exploited so far. At the moment, pre-clinical HCD are almost entirely considered in order 
to contextualize the findings in the index 
study as required by the majority of the test guidelines (Dertinger et al 2022, Kluxen et al 2021, Menssen 2023),
either as a quality check for the CCG (Menssen and Schaarschmidt 2019, Menssen and Schaarschmidt 2022,
Menssen et al. 2024, Menssen and Rathjens 2025, Greim et al 2003, EFSA 2025, Valverde-Garcia 2019) or
in order to evaluate the effect sizes found in the treated cohorts (Gart et al. 1979, EFSA 2017, Zarn et al 2024). 

Current projects, such as \emph{etransafe} (Sanz et al. 2023)
or \emph{vict3r} (Steger-Hartmann et al. 2025) focus on the development of so called virtual control groups (VCG). 
This concept was introduced in 2020, and aims to augment or even completely replace 
the CCG by historical observations drawn from a shared data repository (Steger-Hartmann et al. 2020). 
Since then, several authors worked on different aspects of VCG generation, mainly in the context of
acute toxicity studies and continuous endpoints (Gurjanov et al. 2023. 
Gurjanov et al. 2024, Sato et al. 2026, Sato et al. 2024).

An alternative approach towards the reduction of animal use in the CCG are Bayesian borrowing approaches,
which are more commonly applied in clinical studies, but seem to be almost entirely overlooked 
in the current discussion about
 the reduction of animal usage. Given the fact that seminal papers on Bayesian borrowing
were written in the early 1980s in the context of long-term carcinogenicity studies (Tarone 1982a, Smythe et al. 1986)
and in vitro cell assays (Tarone 1982b) and methodological development continued until the 2010s (Kitsche et al. 2012),
this is surprising!
 
In contrast to non-clinical toxicology, Bayesian borrowing has received much greater attention 
in the context of clinical trials. Consequently, most of the key concepts for these approaches
such as \emph{power priors} (Ibrahim and Chen 2000), \emph{commensurate priors} (Hobbs et al. 2011) or 
\emph{meta-analytic predictive (MAP)} prior distributions (Schmidli et al. 2014) emerged from 
medical applications. Several theoretical aspects for borrowing as well as examples for its application
in clinical trials can be found in literature. For example, Wang et al. 2022 proposed an approach that combines 
 propensity-score matching of historical and current patients with Bayesian borrowing.  
An approach for the translation of pre-clinical information into phase~1 oncology trials
was proposed by Zheng and Hampson 2019. 

Clinical real life applications of Bayesian borrowing and HCD use can be found for example in 
Chen et al. 2024 who published results of a phase~3 trial on asthma medication that used a design in 
which information from a predecessor trial on the same endpoint was borrowed. Richeldi et al. 2022 published
results from a phase~2 trial on idiopathic pulmonary fibrosis that was based on a 2:1 randomization scheme
in which the CCG was augmented via a MAP~prior, whereas Wang et al. 2023 provide an overview about EMA 
approvals for oncology trials that used external control arms. 

Software implementations for the MAP approach are publicly available from
the R~package \texttt{RBesT} (Weber et al. 2021). This R package provides functionality to apply the MAP approach
to different scales of possible endpoints. Furthermore it provides methodology for design evaluation 
in order to derive the type-1 error and power for different borrowing scenarios for two-arm studies. Applications
of the commensurate prior (combined with propensity score matching) are implemented in the R~package 
\texttt{psborrow2} (Secrest and Gravestock 2025).

Given the vast amount of literature available on Bayesian borrowing in clinical settings, 
one of the main purposes of this manuscript is to draw attention towards its possible applications 
in non-clinical toxicological studies. It has to be noted that one of the major differences in 
the experimental design between clinical and non-clinical studies is the fact that 
clinical studies are mainly comprised of an untreated control and one treatment arm,
whereas non-clinical studies are usually conducted based on several cohorts which are treated with different 
dosages of the compound of interest and which are compared against an untreated control. This 
adds a further level of complexity: the problem of simultaneous inference has to be accounted for 
by adjustment.

In the following sections we will demonstrate how HCD can be incorporated in the planing and conduct
of long-term carcinogenicity studies. Two approaches for incorporation of HCD are proposed: 
An empirical Bayes approach in which the estimates for model parameters are directly 
drawn from the HCD, as well as an approach in which the HCD is accounted for via a MAP-prior.
Furthermore, we will show how simultaneous credible limits (and intervals) 
for risk-ratios can be derived using the approach of Besag et al. 1995.
Note that the entire implementation of the proposed approaches is based on existing 
concepts (adjustment following Besag et al. 1995) and R~packages \texttt{RBesT} (Weber et al. 2021), 
\texttt{predint} (Menssen 2025), \texttt{BSagri} (Schaarschmidt 2018) and \texttt{mratios} (Djira et al. 2025) and 
is therefore relatively simple to apply in practical toxicological routine.



\section{Methods}


\subsection{Beta-binomial model} \label{sec::BB_model}
The hierarchical nature of the historical control data gives rise to systematic between-study variability, and it is natural to model dichotomous endpoints based on the \emph{beta-binomial 
distribution}. In this model, the
number of successes within each control group is binomial
\begin{equation*}
	Y_h \sim \binomdistn(\pi_h, n_h)
\end{equation*}
with $n_h$ as the number of experimental units within each control group. The control groups success probabilities $\pi_h$ follow a beta-distribution
%
\begin{equation}
	\pi_h \sim \betadistn(a, b) \label{eq::beta}
\end{equation}
with $\expect(\pi_h)=\pi=a/(a+b)$ and $\var(\pi_h)=\sigma^2 = \frac{ab}{(a+b)^2 (a+b+1)}$.
Furthermore, the overall expected number of successes is $\expect(Y_h)=n_h \pi$ and
$\var(Y_h)=n_h \pi (1-\pi)\big [ 1 + (n_h-1)\rho \big]$. The intra-class correlation 
coefficient is given by $\rho=1/(1+a+b)$.

The success probability of a further control group, $\pi_0$, with expectation
\begin{equation}
	\expect(\pi_0)= \delta \pi = a_0 / (a_0 + b_0) \label{eq::delta}
\end{equation} 
can also be assumed to be beta distributed
\begin{equation*}
	\pi_0 \sim \betadistn(a_0, b_0)
\end{equation*}
with $\delta$ denoting a possible shift between the expected success probabilities
in the  historical and the current control groups.

If the current trial is comprised of $(M+1)$~groups (with $m=0$ denoting
the control and $m=1,\ldots,M$ denoting the treatment groups), the observed numbers within
each group are again binomial with
\begin{gather}
	Y_0 \sim \binomdistn(\pi_0, n_0) \\ 
	Y_m \sim \binomdistn(\Delta_m \pi_0, n_m) \label{eq::Ym}
\end{gather}
with $\Delta_m$ denoting 
possible treatment effects in the concurrent study relative to the common baseline~$\pi_0$.


\subsection{Borrowing using hierarchical models} \label{sec::bayes_borrowing}

Leveraging historical control data to borrow information for the 
current trial is based on the assumption of \emph{exchangeability} between 
the historical and current controls. This means that $\delta=0$ in equation (\ref{eq::delta}). 
Consequently, it is explicitly assumed that the historical and current controls 
originate from a \emph{common} data generating process, 
such that equation~(\ref{eq::beta}) holds for $h=0,1,\ldots,H$ (including $h=0$).

Within the context of the binomial model, use of a (conjugate) beta prior is computationally convenient, as it subsequently allows for some analytic computations.
Based on the exchangeability assumption with prior $\pi_0\sim\betadistn(a,b)$, the posterior (for $y_0$~observed events) becomes
\begin{equation*}
	p(\pi_0 | y_0) = \betadistn(a + y_0, b + n_0-y_0).
\end{equation*}
The (true) population parameters~$a$ and~$b$, however, are not known; they can only be learned to a certain degree from the historical data, and prior information may again be expressed (approximated) via a beta distribution.

An alternative way for prior specification is the meta-analytic predictive (MAP) prior
approach following Schmidli et al. 2014.
In contrast to the beta-binomial model outlined above, this approach was developed within a \emph{normal-normal hierarchical model (NNHM)} framework that is commonly utilized in meta-analysis applications, and which may be applied here based on logit-transformed probabilities (or \emph{log-odds}) $\theta_h = g(\pi_h) = \log(\pi_h / (1-\pi_h))$.
On the transformed scale, a normal distribution is assumed
\begin{equation}
	\theta_0, \theta_h \sim \normaldistn(\mu, \tau^2) \label{eq::normal_model}
\end{equation}
with population mean~$\mu$ and between-study variance~$\tau^2$. Priors for
the hyper-parameters~$\mu$ and~$\tau$ are commonly specified independently, so that $p(\mu, \tau) = p(\mu) \times p(\tau)$. Further information
on the prior settings is given in section~\ref{sec::comp_det} on computational details below. 
Based on the model specified in~(\ref{eq::normal_model}) and the hyper-prior, the
MAP is derived and Markov chain Monte Carlo (MCMC) sampling can be used to derive a 
sample $\theta_0^1, \ldots, \theta_0^S$ on the logit scale. Following Schmidli et al. 2014,
the back-transformed MCMC sample
$\pi_0^s=g^{-1}(\theta_0^s)$ can be approximated by a beta mixture
\begin{equation}
	p(\pi_0 | y_h) = \sum_{k=1}^K \omega_k \betadistn(a_k, b_k) \label{eq::MAP}
\end{equation}
The Beta-mixture is estimated using the expectation-maximization algorithm and the 
ap\-pro\-pri\-ate number of components is chosen based on Akaike Information Criterion 
(AIC) selection as described by 
Weber et al. 2021.

Use of a beta mixture with a small number~$K$ of components leads to model specifications that are relatively simple, as well as easily communicated and implemented. Also, conjugacy of the beta distribution simplifies matters in case of a beta mixture prior;
the posterior then results as
\begin{equation*}
	p(\pi_0 | y_0, y_h) = \sum_{k=1}^K \tilde{\omega}_k \betadistn(a_k + y_0, b_k + n_0-y_0)
\end{equation*}
that is, again as a beta mixture; the weights $\tilde{\omega}_k$ are also updated based on the fit (marginal likelihood) of each mixture component to the data.

Taking HCD into account requires a careful selection of historical controls that 
are comparable to the current one. However, in practice it is often unclear whether the 
restrictive assumption of exchangeability is appropriate. Therefore, it is recommended to anticipate potential non-exchangeability in the analysis model and
\emph{robustify} the derived prior with an uninformative or vague component, such that
\begin{equation*}
	p(\pi_0 | y_h)^{rob} = (1-\omega^{rob}) p(\pi_0 | y_h) + \omega^{rob} p(\pi_0)^{u}.
\end{equation*}
Note that the two components in the robustified mixture prior can be interpreted as two competing
models: The first suggests borrowing from HCD, whereas the second suggest to fully ignore
the historical information. Therefore, the weight for the uninformative component $\omega^{rob}$ 
specifies the magnitude of ''skepticism'' against the prior distribution and directly relates
to the probability of a possible data-prior conflict (Röver et al. 2018). 
Robustification of the simple beta prior from equation~(\ref{eq::beta}) yields
\begin{equation*}
	p(\pi_0 | y_h)^{rob} = (1-\omega^{rob})  \betadistn(a, b) + \omega^{rob}  \betadistn(1, 1),
\end{equation*}
assuming that a $\betadistn(1,1)$ distribution (a uniform distribution) represented an uninformative prior; 
the posterior then becomes
\begin{equation*}
	p(\pi_0 | y_0, y_h)^{rob} = (1-\tilde{\omega}^{rob})  \betadistn(a+y_0, b+ n_0 - y_0) + \tilde{\omega}^{rob}  \betadistn(1+ y_0, 1+ n_0 - y_0).
\end{equation*}
Similarly, robustification of the MAP prior given in equation~(\ref{eq::MAP}) yields
\begin{equation*}
	p(\pi_0 | y_h)^{rob} = (1-\omega^{rob})  \sum_{k=1}^K \omega_k \betadistn(a_k, b_k) + \omega^{rob}  \betadistn(1, 1)
\end{equation*}
(essentially adding an additional $K+1$th component to the mixture),
and the posterior becomes
\begin{equation*}
	p(\pi_0 | y_0, y_h)^{rob} = (1-\tilde{\omega}^{rob}) \sum_{k=1}^K \tilde{\omega}_k \betadistn(a_k + y_0, b_k + n_0 - y_0) + \tilde{\omega}^{rob}  \betadistn(1+  y_0, 1+ n_0 - y_0) 
\end{equation*}
Note that the posterior weights for each informative component $\tilde{\omega}_k$ as well as the weight for
the uninformative component  $\tilde{\omega}^{rob}$ again are updated depending on the posterior probabilities 
for the two main components of the distribution (uninformative $\betadistn(1,1)$ and informative MAP prior).

This shows that robustification yields a posterior distribution which dynamically borrows historical
information on two stages: First, because the variance of the (informative) prior distribution is proportional to the 
between-study variance, the informativeness of the prior decreases with increasing between-study variance. Second,
because the posterior weights directly depend on posterior probabilities for the two components (uninformative, 
informative), robustification can be interpreted as model averaging that automatically accounts for possible 
drift.

\subsection{Common-effect borrowing} \label{sec::freq_borrowing}
 
In its simplest form, borrowing may be implemented as complete pooling of historical and current 
control data, such that $y^{pool}_0=y_0 + \sum_1^{H} y_h$ and $n^{pool}_0= n_0 + \sum_1^{H} n_h$.
The estimate for the underlying common binomial proportion is 
\begin{equation}
	\hat{\pi}^{pool}_0 =  \omega_0 \hat{\pi}_0 + \omega \hat{\pi}
\end{equation}
with $\omega_0 = \frac{n_0}{n^{pool}_0}$, $\hat{\pi}_0 = y_0 /n_0$,
$\omega = \frac{\sum_1^{H} n_h}{n^{pool}_0} $ and $\hat{\pi} = \frac{\sum_h y_h}{\sum_h n_h}$. 
Assuming independence between~$\hat{\pi}_0$ and~$\hat{\pi}$, the variance of~$\hat{\pi}^{pool}_0$
is given by
\begin{equation}
\var(	\hat{\pi}^{pool}_0) = \omega_0^2 \var(\hat{\pi}_0) + \omega^2 \var(\hat{\pi}) \label{eq::var_pool}
\end{equation}
with $\var(\hat{\pi}_0)=\frac{\pi_0(1-\pi_0)}{n_0}$ and $\var(\hat{\pi}) = \frac{\pi (1-\pi)}{\sum_h n_h}$.
It has to be noted, that such complete pooling severely depends on the assumption that CCG and HCD are independent
realizations from \emph{the same} binomial distribution ($y_h \stackrel{\text{iid.}}{\sim} \binomdistn(n_h, \pi)$) and therefore
ignores possible between-study variation. In other words, for many practically relevant settings,
$\var(\hat{\pi}^{pool}_0)$ reflects some degree of model-misspecification that leads to an underestimation
of the true variability of the underlying data generating process.

An obvious alternative to complete pooling might be a \emph{test-then-pool} approach. Such an approach depends on a
significance test between the success rates of the concurrent control and each historical control group;
only historical controls that do not significantly differ from the current controls are pooled 
with the current control (Viele et al 2014). 


\subsection{Effects of borrowing on the estimation of success probabilities} \label{sec::shrink}


If historical information is considered, the estimate for the current success probability~($\pi_0$) is systematically drawn towards the overall historical success rate $\expect(\pi_h)=\pi$ (the expected value for the beta distribution in~(\ref{eq::beta})). 
Such \emph{shrinkage} is largest for a common-effect approach,
while less shrinkage will take place for models allowing for heterogeneity.
Analogously, the different estimates' variances are affected; an estimate considering the current data in isolation has larger variance, while consideration of additional data generally reduces variance of an estimate according to the amount of shrinkage.

For a frequentist common-effect estimator,
this is directly visible from equation~(\ref{eq::var_pool}), in which both the weights and the
variances depend on the sample size in the current control and the HCD\@. With an increasing
number of experimental units in the HCD, the variance for the historical average success probability
$\var(\hat{\pi})$ is up-weighted and hence plays a dominant role. However, the variance for the 
historical success probability decreases with an increasing number of historical experimental units.
Because the number of experimental units in the current control group $n_0$ is fixed to a relatively small
number, practically, $\var(\hat{\pi}_0)$  drops out and $\var(\hat{\pi}_0^{pool})$ goes towards zero, 
if the number of historical experimental units goes towards infinity. 

Within a Bayesian framework, \emph{complete pooling} occurs if (as in Section~\ref{sec::freq_borrowing}) a common underlying binomial probability is assumed for historical and current data. In that case, the posterior becomes
\begin{equation}
	p(\pi_0 | y_0, y_h)=\betadistn\Bigl(y_0 + \sum_h y_h,\, (n_0 - y_0) + (\sum_h n_h - \sum_h y_h)\Bigr).
\end{equation}
On the other hand, if historical information is taken to be irrelevant and no information is borrowed,
the posterior becomes
\begin{equation}
	p(\pi_0 | y_0, y_h)=\betadistn(y_0 + 1, (n_0 - y_0) + 1).
\end{equation}
In practical application to real life HCD, the estimates for the parameters of the beta prior's parameters ($a$, $b$) increase with decreasing
between-study-variability. Therefore, the posterior variance should be relatively close to
the (frequentist) estimate for the variance of the pooled success rate $\widehat{\var}(\hat{\pi}_0^{pool})$,
if between-study variability is practically absent. 
With increasing between-study variability, the estimates for the parameters of the beta prior
decrease, meaning that, practically, the posterior variance will be close to the estimate for the variance of
the success rate in the current control $\widehat{\var}(\hat{\pi}_0)$.


\subsection{Effective sample sizes} \label{sec::ess}

In common-effect pooling approaches, the number of animals from which information is borrowed directly relates to the number of current observations, making it easy to quantify the amounts of information contributed by both sources. 
As soon as between-study heterogeneity is considered (in a hierarchical random-effects approach) this simple correspondence is lost, as the amount of information that is ``borrowed'' from the HCD \emph{decreases} with increasing between-study variability.
In the simple case of a binomial model with a conjugate beta prior, the prior's \emph{effective sample size (ESS)} simply results as~$a+b$. 

However, for mixture priors, the calculation of 
the effective sample size is not straightforward and different concepts have been discussed 
in the literature (see Neuenschwander et al. 2020 and discussions within). In the following,
we will use the ESS based on the \emph{expected local-information-ratio (ELIR)} method for the calculation of effective sample sizes, as rec\-om\-mended by Neuenschwander et al. 2020. 
Analogously to the prior ESS, the \emph{posterior ESS} provides information on an ``effective'' number of experimental units according the posterior's precision.
The (prior) ESS\textsubscript{ELIR} is defined such that the expected posterior ESS equals the prior ESS plus the actual sample size. Robustification of a prior will generally lead to a smaller prior ESS, while the posterior ESS will decrease with increasing prior-data conflict.


\subsection{Simultaneous inference using prior information}

In a typical toxicological study, the outcome of the untreated (or negative) control
is compared to several cohorts which are treated with the compound of interest in different 
dosages. Therefore, different types of hypotheses, such as many-to-one comparisons 
(e.g. Dunnett type) or tests for a (linear) trend (e.g. Williams type) can be formulated and are
required by many test guidelines (Dertinger et al. 2023). Because both approaches
are based on several elementary hypotheses that are tested simultaneously, one needs to adjust for multiple testing. 

With respect to risk ratios, the following hypotheses are tested in a Dunnett-type comparison
\begin{equation}
	H_0 : \bigcap_{m=1}^M \pi_m / \pi_0 \leq 1 \text{ vs. } H_1 : \bigcup_{m=1}^M \pi_m / \pi_0 > 1 \label{eq::hypothesis}
\end{equation}
and the corresponding lower confidence (or credible) limits need to control the \emph{familywise error rate (FWER)}
\begin{gather}
	FWER = 1 - P(l_m \leq 1 \text{ } \forall \text{ } m=1, \ldots, M) = \alpha
\end{gather}
rather than the individual type-1-error $1 - P(l_m \leq 1)$ for each of the $M$ elementary hypotheses.

The compound of interest may be claimed 
to be potentially harmful, if at least one difference between the treatment groups and the control
is found to be significant. Therefore, we also assessed the power to detect at least one increase in one of the treatment 
groups over the control (\emph{any-pair-power, APP})
\begin{equation}
	APP = P(1 < l_m  \text{ } \exists \text{ } m=1, \ldots, M).
\end{equation}

Under the Bayesian framework one can derive the posterior for the CCG based on an informative prior as shown above.
However, since historical knowledge on the compound of interest is usually not available in practice,
uninformative priors need to be applied to the treatment groups; using a uniform ($\betadistn(1,1)$) prior, the posterior becomes 
$\betadistn(1 + y_m, 1 + n_m-y_m)$.

It is possible to construct simultaneous, one-sided ($1-\alpha$) credibility 
sets for the $M$ ratios of interest, $\boldsymbol{\Delta} = ( \pi_{1}/\pi_{0}, \pi_{2}/\pi_{0}, ..., \pi_{M}/
\pi_{0})$, based on $B$ samples from the posterior of $(\pi_0,...,\pi_M)$. Because only lower limits 
are of interest in this application, the method is a special case of the methods described in Mandel and 
Betensky (2008), or Besag et al. (1995). The corresponding algorithm is given in the box below.

As a frequentist comparator for simultaneous confidence limits for risk ratios, 
we consider the approach of Hothorn et al. 2008 which --- for binomial endpoints --- is based on 
a multivariate normal distribution that recognizes the correlation
structure of the different tested hypotheses. If HCD are considered 
via pooling, the only difference to an approach that ignores the historical
information is that the confidence limits for $\pi_m / \pi_0$ depend on
$\hat{\pi}_0^{pool}$ and $\hat{\pi}_m$ (and their standard errors) rather than 
on $\hat{\pi}_0$ and $\hat{\pi}_m$ (with $\hat{\pi}_m$ as the estimates for the success 
probabilities in the treatment groups).

\begin{mdframed}

\begin{enumerate}
	\item Draw $B$ samples from the posterior of $(\pi_{0}, \pi_{1},..,\pi_{M})$, with index $b=1,...,B$ 
	denoting the index of samples, and $(\pi_{0b}, \pi_{1b},..,\pi_{Mb})$ the sampled values.

	\item For each sampled vector $b$, the ratios of interest need to be computed
	$\boldsymbol{\Delta_{b}} = ( \pi_{1b}/\pi_{0b}, \pi_{2b}/\pi_{0b}, ..., \pi_{Mb}/\pi_{0b})$. 
	This results in a sample of $B$ vectors which can be stored in a 
	$(B \times M)$ matrix in which rows correspond to samples from posterior, 
	and columns correspond to the ratios of interest.

	\item Rank the values in the columns ($m=1,...,M$) of the sample matrix, separately for each column. 
	This results in the ranks $r_{mb}$, and the corresponding order statistics $\Delta_{m}^{(b)}$.

	\item Within each row ($b = 1,...,B$), compute $R_{b} = B + 1 - \min_{m=1,...,M}(r_{mb})$.

	\item Order $R_{b}$ across $b=1,...,B$ to obtain its order statistics $R^{(b)}$.

	\item Let $q$ denote the nearest integer to $B(1-\alpha)$. 

	\item Select the $q$th value from these order statistics, $R^{(b)}$, i.e. $R^{(q)}$

	\item The $M$ lower credibility limits $l_m$, for the $m=1,...,M$ ratios of interest 
	can now be obtained by selecting the $(B+1-R^{(q)})$th order statistics from each columns 
	order statistics $\Delta_{m}^{(b)}$: $l_m = \Delta_{m}^{(B+1-R^{(q)})}$.
\end{enumerate}

\end{mdframed}

\section{Implementation and computational details} \label{sec::comp_det}

All methods as well as the simulation studies were implemented in R (R Core Team 2025).

\subsection{Bayesian approaches}
In order to keep the approach as simple as possible, the parameters of the beta prior 
for~$\pi_0$
were directly estimated
from the HCD\@. The estimation of $\hat{a}$ and $\hat{b}$ relies on methods of moments 
estimates for the success probability and the intra-class correlation 
(following Lui et al 2000) as implemented in the \texttt{pi\_rho\_est()} function 
of the R package \texttt{predint} (Menssen 2025). The parameters were derived as follows
\begin{gather*}
	(\hat{a}+\hat{b}) = \frac{1-\hat{\rho}}{\hat{\rho}} \\
	\hat{a} = \hat{\pi} (\hat{a}+\hat{b})\\
	\hat{b} = \hat{a} - (\hat{a}+\hat{b})
\end{gather*}
However, the estimate for the intra-class correlation can become negative, indicating empirical underdispersion which is not plausible. 
For this case, Menssen and Schaarschmidt 2019 proposed its restriction to $\max(0.00001, \hat{\rho})$. 
In the (apparent) absence of between-study variation, this approach leads to inflated estimates for the 
parameters of the beta distribution. Therefore, we further restrict both estimates to $\min(\hat{a}, \sum_h y_h)$
and $\min(\hat{b}, \sum_h (n_h - y_h))$. Because $\hat{a}$ and $\hat{b}$ were directly estimated
from the HCD, we will refer to this methods as the \emph{empirical Bayes} approach in the
remainder of the manuscript.

Estimation of the MAP prior was done using the implementation in the \texttt{RBesT} package (Weber et al. 2021).
The priors for the hyper-parameters are defined as follows: 
The prior for the mean~$\mu$ is $\normaldistn(0, 2^2)$, which is the default in \texttt{RBesT}. Following the
recommendations in the vignette of \texttt{RBesT} (Weber et al. 2025), the prior for 
the between-study standard deviation~$\tau$ was set to $\halfnormaldistn(0,1)$ in cases
in which the binomial success probabilities ranged between 0.2 and 0.8. If the 
binomial success probabilities were below 0.2, the standard deviation of the half-normal prior was derived based on the $n^\infty$ approach of Neuenschwander et al. 2010.
Based on the sampling standard deviation
\begin{gather*}
	s = \frac{1}{\sqrt{\bar{\pi} (1-\bar{\pi})}} \quad \text{ with} \quad
	\bar{\pi} = \frac{1}{H}\sum_h \frac{y_h}{n_h}
\end{gather*}
a conservative prior for $\tau$ is given by $p(\tau) = \halfnormaldistn(0, s/2)$.

\subsection{Frequentist approaches}

In a first step, all frequentist methods proposed above were implemented depending
on simple generalized linear models (GLM) fit on log-link using \texttt{stats::glm()}. 
Note that instead of the canonical logit-link, the log-link was used for model fitting, 
meaning that the desired lower confidence limits were formulated for differences between 
$log(\pi)_m$ and $log(\pi)_0$, such that back-transformation to response-scale 
yields confidence limits for the desired \emph{risk ratios (RR)}.

Because the simple GLM is prone to the Hauck-Donner effect (massive inflation of the 
standard error in case of zero findings and affected lower limits close to zero),
the frequentist approaches were also implemented
based on a Bayesian version of a GLM using \texttt{arm::bayesglm()} (Gelman and Yu-Sung 2024) 
using their non-informative default priors for the hyperparameters. Note that the Bayesian 
generalized linear model (BGLM) 
was interpreted in a frequentist sense and was only applied to reduce the impact of the 
Hauck-Donner effect (due to the regularizing effect of the hyperpriors, the standard error is far less inflated than 
in the simple GLM). 

For all frequentist approaches, simultaneous lower confidence limits were computed following 
Hothorn et al. 2008 using the \texttt{emmeans} package (Lenth and Piaskowski 2025)
with the option \texttt{adjust="mvt"}. For the test-then-pool approach,
the binomial proportion of each historical control group was tested pointwise against 
the one from the current control ($\alpha=0.05$). Only historical controls that do not 
differ significantly from the current control were used for pooling.


\section{Simulations}

The aim of the simulation was an assessment of the FWER and 
power of the proposed methods and their non-borrowing counterparts outlined in Table~\ref{tab::methods}. It was of special interest to evaluate the effect of a severe reduction
of animals in the control group in order to check for a possible reduction of animal use in 
long-term carcinogenicity studies.

All test-procedures were evaluated based on the Dunnett-type hypothesis 
given above in Equation~(\ref{eq::hypothesis}).
Most of the values for model parameters used for Monte Carlo simulations 
(Table~\ref{tab::ltc_setting}) reflect the properties of real life HCD of two-year carcinogenicity 
studies obtained from the historical control data base of the U.S. National 
Toxicology Program (Menssen and Schaarschmidt 2019).

\begin{table}[H]
\caption{Overview about methods}
\begin{tabular}{lllll}
 \textbf{Label} & \textbf{Description} & \textbf{Type} & \textbf{Borrowing}  \\ \hline
 GLM & \shortstack{Current data only (GLM) + \\ Hothorn et al. 2008} & Frequentist &  No  \\ \hline
 BGLM & \shortstack{Current data only (Bayesian GLM) + \\ Hothorn et al. 2008} & Frequentist &  No  \\ \hline
 Naiv pool. & \shortstack{Naiv pooling (GLM) + \\ Hothorn et al. 2008} & Frequentist & Yes   \\ \hline
 TAP & \shortstack{Test HCD before pooling (GLM) + \\ Hothorn et al. 2008} & Frequentist &  Yes \\ \hline
 Naiv pool. (B) & \shortstack{Naiv pooling (Bayesian GLM) + \\ Hothorn et al. 2008} & Frequentist & Yes   \\ \hline
 TAP (B) & \shortstack{Test HCD before pooling (Bayesian GLM) + \\ Hothorn et al. 2008} & Frequentist &  Yes \\ \hline
 $\betadistn(1,1)$ & \shortstack{Uninformative prior + \\ Besag et al 1995} & Bayesian & No   \\ \hline
 Emp. Bayes & \shortstack{Beta prior with MoM-estimates + \\ Besag et al 1995} & Bayesian & Yes   \\ \hline
 Emp. Bayes (robust)& \shortstack{Robustified Beta prior with MoM-estimates + \\ Besag et al 1995} & Bayesian & Yes   \\ \hline
 MAP & \shortstack{Meta-analytic predictive prior + \\ Besag et al 1995} & Bayesian &  Yes  \\ \hline
 MAP (robust) & \shortstack{Robustified meta-analytic predictive prior + \\ Besag et al 1995} & Bayesian & Yes \\ \hline
\end{tabular} 
\label{tab::methods}
\end{table}

\begin{table}[h!]
\centering
\caption{Parameter values used for the  simulation} 
\begin{tabular}{llc}
\hline 
\textbf{Parameter} & \textbf{Notation} & \textbf{Values} \\ 
\hline 
No. of historical studies & $H$ & \textbf{5},  \textbf{10},  \textbf{20},  100 \\ 

Binomial probability ($\frac{a}{a+b}$)& $\pi$ & \textbf{0.01},  \textbf{0.1},  \textbf{0.2}, \textbf{0.3}, \textbf{0.4}, \textbf{0.5} \\ 

Intra-class correlation ($\frac{1}{1+a+b}$)\textsuperscript{$\ast$} & $\rho$ & \textbf{1e-05}, \textbf{0.01}, \textbf{0.04}, 0.08 \\ 

Historical cluster size & $n_h$ & \textbf{50} \\ 

No. of groups in current trial & $M$ & 4 \\ 

Cluster size of CCG & $n_0$ & 10, \textbf{50} \\

Cluster size of TRG & $n_m$ &  \textbf{50} \\ 

Shift between HCD and CCG & $E(\pi_h) / E(\pi_0) = \delta$ &  1, 1.25, 1.5 \\ 

Max. increase over the control & $\Delta_M$ & 1.25, 1.5, 1.75 \\
\hline 
\end{tabular} \\ [0.5ex]
\raggedright
\scriptsize
\textbf{Bold numbers:} Parameter values that reflect real life data in the NTP historical control data base. \\
\textsuperscript{$\ast$} Intra-classs correlation ($\rho$) settings correspond to $a+b=\frac{1}{\rho}-1 \in \{99\,999,\, 99,\, 24,\, 11.5\}$. 
\label{tab::ltc_setting}
\end{table}


\subsection{Family wise error rate}

\begin{figure}[H]
\centering
\includegraphics[width=1\columnwidth]{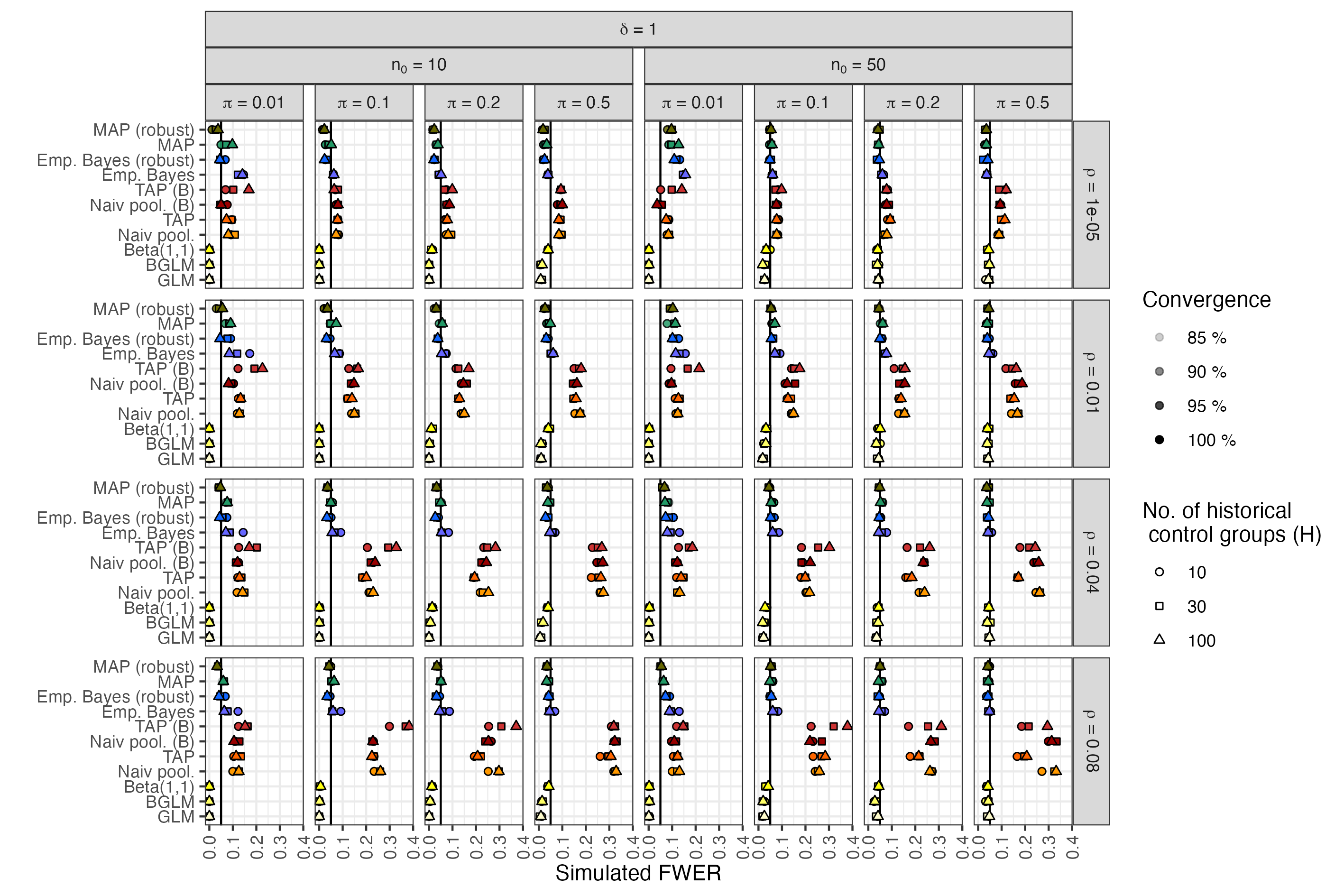}
\caption{Simulated family wise error rate (FWER) in the absence of a data-prior conflict ($\expect(\pi_h) = \expect^(\pi_0) = \pi$).}
\label{fig::FWER}
\end{figure}

\begin{figure}[H]
\centering
\includegraphics[height=0.9\textheight]{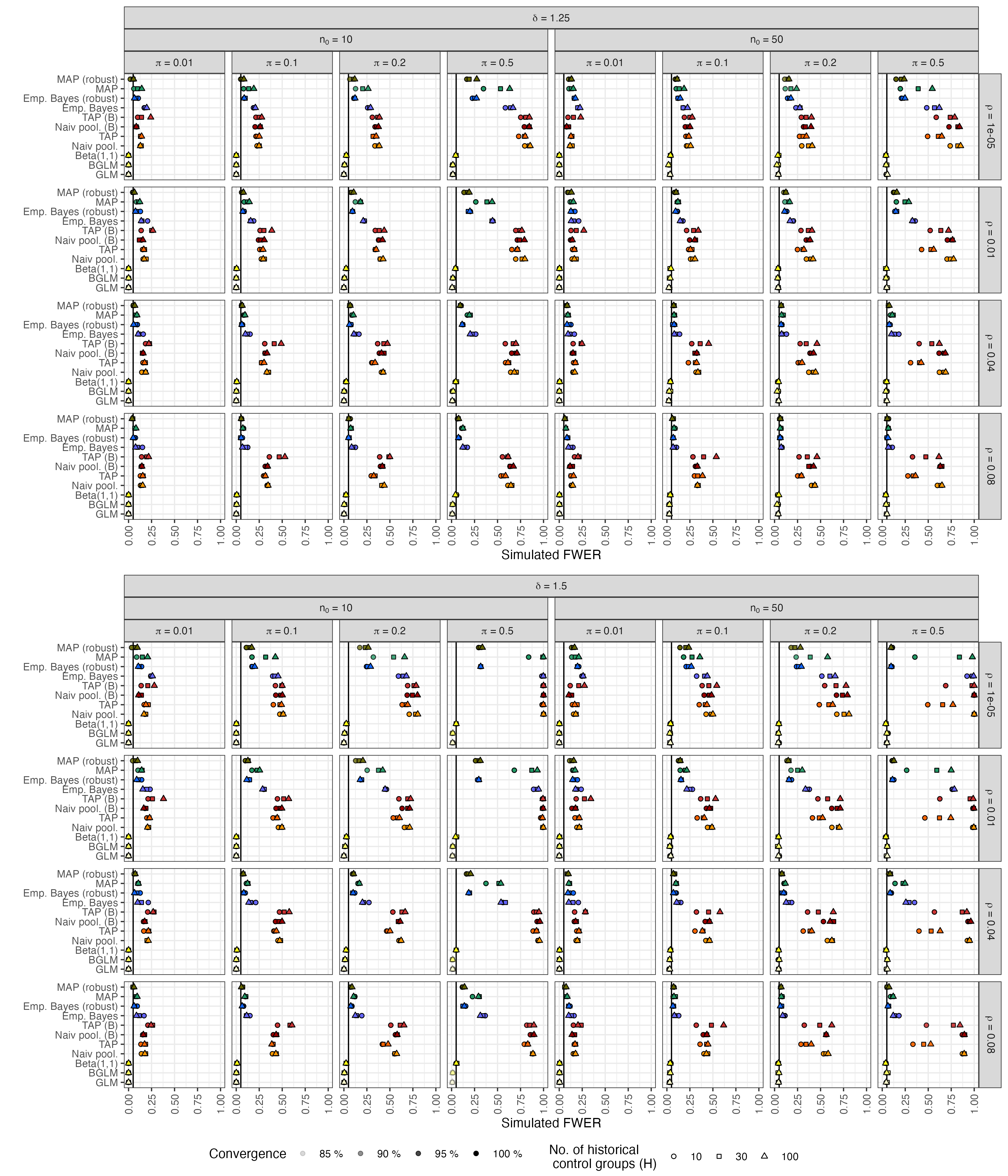}
\caption{Simulated FWER in the presence of a data-prior conflict. 
Upper panel: moderate drift ($\expect(\pi_0) = \delta \expect(\pi_h) = 1.25 \pi$).
Lower panel: large drift ($\expect(\pi_0) = \delta \expect(\pi_h) = 1.50 \pi$).}
\label{fig::FWER_drift}
\end{figure}

In order to sample data under~$H_0$, the maximum increase over the control  
was fixed to~$\Delta_M=1$. For each of the remaining 288~combinations of parameter 
values (Table~\ref{tab::ltc_setting}), $s=1, \ldots, 2000$ data sets were sampled from 
the beta-binomial model described in Section~\ref{sec::BB_model}. 
Simultaneous lower confidence or credible limits $l_m^s$ were calculated for each of the sampled
data sets and the FWER was estimated to be 
\begin{gather*}
	\widehat{FWER}=1 - \frac{\sum_s^S I_s}{S} \text{ with} \\
	I_s = 1 \text{ if } l_m^s \leq 1 \text{ } \forall m=1, \ldots, M  \\
	I_s = 0 \text{ if } 1 < l_m^s \text{ } \exists m=1, \ldots, M 
\end{gather*}
The results of the simulation are given in Figures~\ref{fig::FWER} and~\ref{fig::FWER_drift}. Note that,
due to a lack of convergence, the MAP prior could not be derived for all of 
the sampled data sets, especially if $\pi=0.01$ and rising intra-class-correlation. 
Also, both types of GLM did not reach convergence in all cases. 
In these cases, the FWER was calculated based on the number of sampled data sets for which the model did converge. 

%

Figure~\ref{fig::FWER} shows the simulated FWER in the absence of a data-prior 
conflict ($\delta=1$). If the sample size in the current control remains unchanged ($n_0=50$), the methods that 
do not rely on borrowing (yellow color in Figure~\ref{fig::FWER}) behave in the same fashion: 
For rare events ($\pi=0.01$) they are
heavily conservative, but their simulated FWERs approach the nominal~0.05 with increasing~$\pi$. 
If the sample size in the current control is reduced to $n_0=10$ (but remains at~$n_m=50$ in the 
treatment groups), simultaneous confidence intervals calculated based on both versions of a GLM 
behave heavily conservative, regardless of the success probability~$\pi$. 
Also, simultaneous credible intervals behave heavily conservative if an uninformative $\betadistn(1,1)$ prior 
is applied and the success probability is low, but their 
simulated FWERs reach the nominal level if $\pi$ approaches~0.5.

The simulated FWER of both frequentist borrowing approaches (orange and brown colors in Figure~\ref{fig::FWER}) 
does not reach the nominal level of 0.05, but always remains above.
Except for the test-and-pool approach that is implemented based on a BGLM, the simulated FWER of the 
frequentist borrowing methods depends almost solely on the magnitude of between-study variation (the
intra-class correlation $\rho$) and is not affected by the sample size in the current control $n_0$.

If the success probability $\pi$ is not too small (at least 0.1), both Bayesian approaches
reach the nominal FWER at least approximately (blue and green colors in Figure~\ref{fig::FWER}). 
However, the empirical Bayes approach that depends on a single beta prior
with parameters directly estimated from the HCD tends to 
approach the nominal level from above. Contrary, using a MAP-prior (green color in Figure~\ref{fig::FWER})
results in a slightly conservative behavior. However, for both methods, robustification results in simulated 
FWERs that are closer to the nominal level than their non-robustified counterparts.
If rare events are considered ($\pi=0.01$) and the current control group's sample size is 
decreased to $n_0=10$, both approaches that rely on robustification approach the nominal FWER with
an increase of available historical controls ($H$), regardless of the magnitude of between-study variation.
However, if the current control group's sample size remains unchanged ($n_0=50$), all borrowing approaches 
remain liberal if rare events are considered.

Figure~\ref{fig::FWER_drift} illustrates simulation results in the presence of a data-prior conflict. 
In this scenario the expected value for the control groups' success probabilities 
is increased between the current and historical controls by factors $\delta=1.25$ or $\delta=1.50$. 
As expected, the methods that do not borrow information from the HCD (yellow colors) 
behave in exactly the same fashion as in scenarios without a data-prior conflict. In contrast, borrowing 
leads to an inflation of the FWER\@. However, in most settings the frequentist approaches are more affected 
than their Bayesian counterparts. In all settings, robustification leads to credible intervals that appear 
less liberal than their non-robustified counterparts. The highest FWER for the robustified borrowing 
approaches (approx.~0.3) was detected in the large-drift scenario (lower panel of Figure~\ref{fig::FWER_drift}), 
if the number of animals in the CCG was reduced to~10, the success probability was high ($\pi=0.5$) and the intra-class 
correlation was moderate ($\rho=0.01$). However, in most other drift-scenarios the FWER was far less affected. 
Nevertheless, only in settings in which the between-study variation was high ($\rho=0.08$), robustification 
leads to a satisfactory control of the FWER\@.

\subsection{Power}

Data sampling was performed in a similar fashion as described above, but 
a data-prior conflict between HCD and CCG was avoided ($\delta=1$).
Furthermore, data mimicking the current trial was sampled based on increasing success 
probabilities, such that the sampled observations in the treatment groups were 
realizations of $\binomdistn(\Delta_m \pi_0, n_m)$. For the simulation, $\Delta_m$ was 
equidistantly increasing from~1 to~$\Delta_M$ as given in Table~\ref{tab::ltc_setting}.

Based on $S=2000$ Monte Carlo samples per parameter combination, 
simulated power to detect at least one increase was defined as
\begin{gather*}
	\widehat{APP}=1 - \frac{\sum_s^S I_s}{S} \text{ with} \\
	I_s = 1 \text{ if }  l_m^s > 1 \text{ } \exists m=1, \ldots, M  \\
	I_s = 0 \text{ if } l_m^s \leq 1 \text{ } \forall m=1, \ldots, M 
\end{gather*}
Similarly as for the simulation of the FWER, not all methods reached convergence in all cases. 
Contrary to the FWER simulation, all methods that rely on Bayesian or frequentist GLMs
were heavily affected in the presence of treatment effects 
rising success probabability and intra-class correlation, convergence could be reached for only approx.~65\% 
of simulated data sets.
 
\begin{figure}[H]
\centering
\includegraphics[width=1\columnwidth]{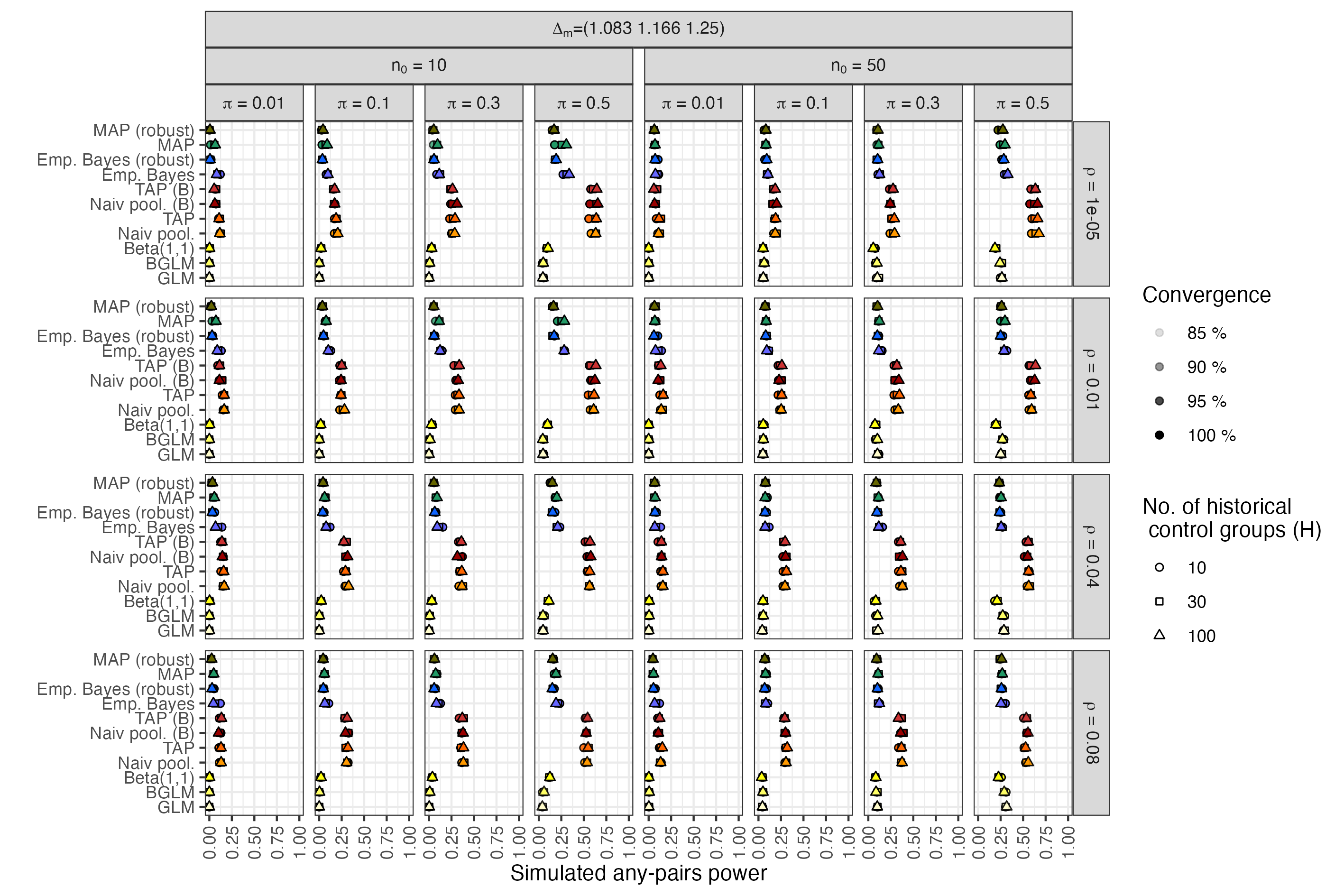}
\caption{Simulated power to detect at least one increase for moderate treatment effects ($\max(\Delta_m)=1.25$).}
\label{fig::power_app_125}
\end{figure}

\begin{figure}[H]
\centering
\includegraphics[width=1\columnwidth]{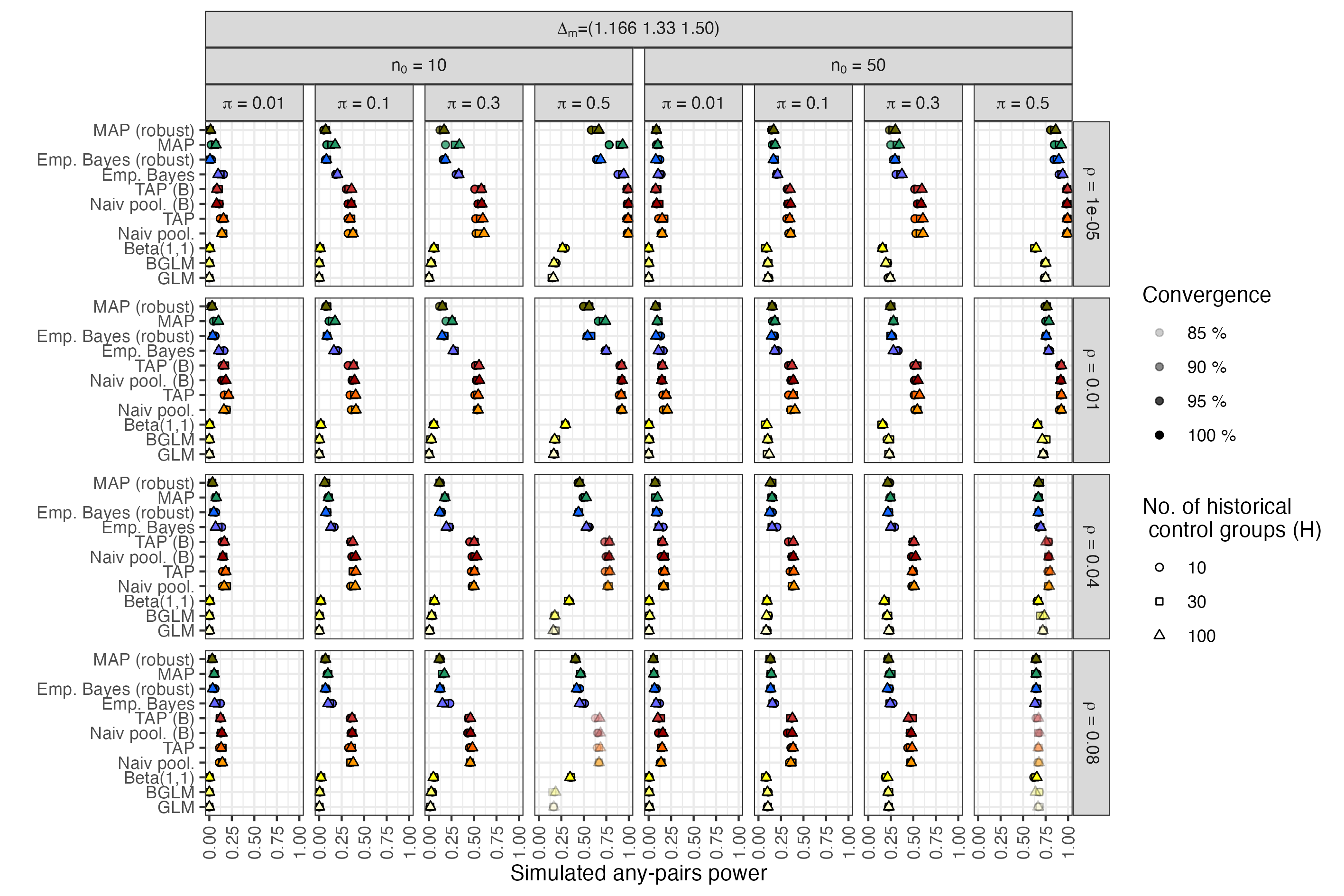}
\caption{Simulated power to detect at least one increase for high treatment effects ($\max(\Delta_m)=1.50$).}
\label{fig::power_app_150}
\end{figure}

\begin{figure}[H]
\centering
\includegraphics[width=1\columnwidth]{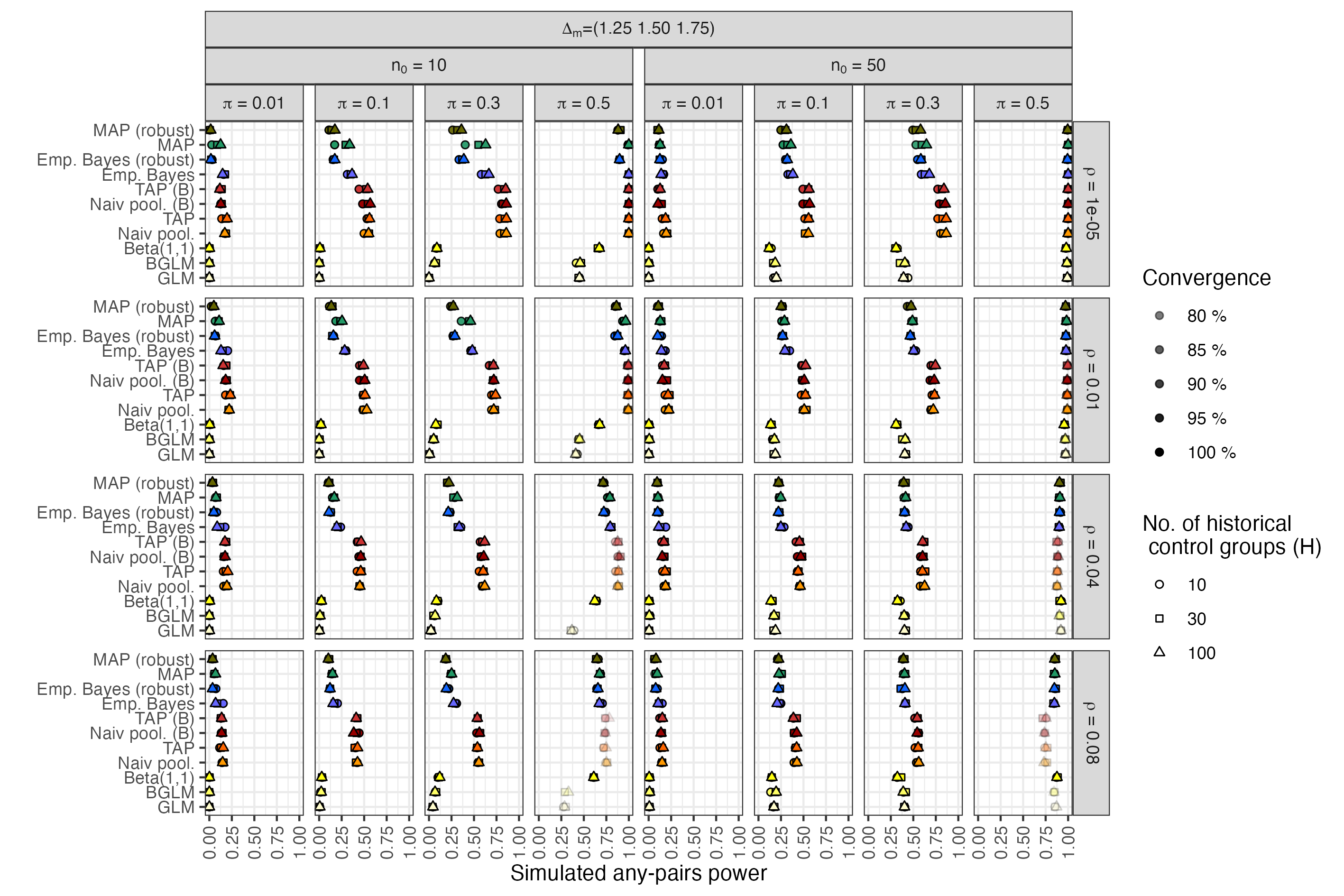}
\caption{Simulated power to detect at least one increase for severe treatment effects ($\max(\Delta_m)=1.75$).}
\label{fig::power_app_175}
\end{figure}

The any-pairs power to detect at least one increase (APP) is presented in Figures~\ref{fig::power_app_125},
 \ref{fig::power_app_150} and~\ref{fig::power_app_175}.
 Generally speaking, long term carcinogenicity studies are heavily underpowered 
if HCD is not taken into account (yellow color in figures). Especially for rare events ($\pi =0.01$ or $\pi=0.1$),
the power to detect an increase in success rates ranges between zero and 0.25.
 However, the Bayesian approach using an 
uninformative prior leads to slightly higher power than the one that is based on a frequentist GLM,
because it is not prone to the Hauck-Donner effect.
In all settings, frequentist pooling (orange and brown colors in the figures) leads to the highest APP gains (up to approx.~80 
percentage points). However, this comes at the cost of heavy FWER inflation.

Bayesian borrowing (blue and green colors in the figures) leads to a moderate gain of the APP, which rises 
with increasing treatment effects (increasing~$\Delta_m$) and with increasing success probability~($\pi$),
while it decreases with increasing between-study variability~($\rho$). A key finding of this study is that
in most scenarios in which the CCG sample size is decreased to~$n_0=10$, but augmented with robustified
priors, the power of the standard design and procedure ($n_0=50$, data modeled based on a GLM) can be retained.
Furthermore, power gains are highest in settings in which the sample size of the control group was reduced to~$n_0=10$.
However, this is mainly driven by the loss of power of the non-borrowing methods compared to the setting in which the sample size of the CCG was unchanged ($n_0=50$).

\section{Applications}

\subsection{Benchmark-comparisons between lower confidence and credible limits based on 18 real-life studies}

A methodological comparison of lower confidence or credible limits was performed on 18
long‑term carcinogenicity studies available in the German Federal Institute for Risk 
Assessment (BfR) database. Study selection followed EFSA core HCD requirements, namely
the same species, strain, testing laboratory, and comparable husbandry conditions (EFSA 2025). 
OECD carcinogenicity and combined chronic/carcinogenicity test guidelines were 
prioritized for their broad regulatory acceptance.

\begin{figure}[H]
\centering
\includegraphics[width=1\columnwidth]{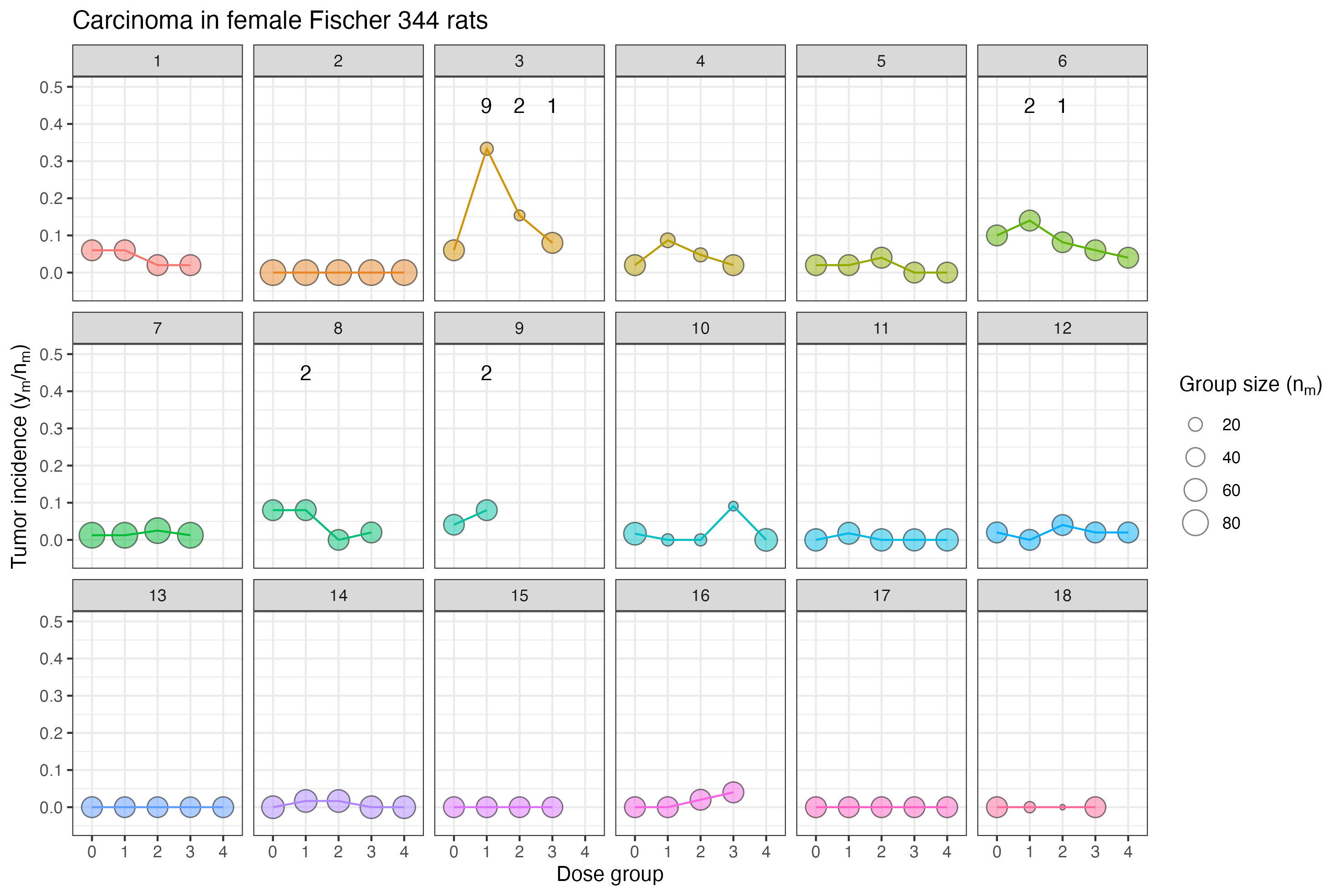}
\caption{C-cell carcinoma incidences in female Fischer~344 rats obtained in the 18 studies provided by the BfR. 
Numbers: Indication for the number of methods that suggest significant increase over control.
9: All methods.
2: Naive pooling and test-and-pool.
1: Test-and-pool.}
\label{fig::BfR_F344}
\end{figure}

Data were extracted from the long‑term groups, which combined animals found dead, euthanised when moribund, 
or sacrificed at the scheduled terminal kill, consistent with standard lifetime tumor incidence reporting.
The data set comprised both male and female animals across a total of 160 dose‑sex groups. The number of treatment 
groups per study per sex ranged from one to four. In 10~studies, four treatment groups were compared against the 
control group, in 7~studies three treatment groups were compared against the control group, and in one study 
only a single treatment group was compared with the control group. In 6~studies, dose levels differed between 
sexes, reflecting sex-specific experimental designs. The numbers of rats examined at the end of the study ranged 
from~8 (in treatment group~2 of study~18) to~80 (all groups in studies~2 and~7), with most groups including 
approximately 50~animals. A total of 247~thyroid c-cell carcinoma cases were recorded across 8,225~animals examined, 
of which 86~cases occurred in female animals. Tumor incidences for thyroid C‑cell carcinomas in female Fischer-344 rats are shown in Figure~\ref{fig::BfR_F344}.

For the analysis of any particular study, the CCGs of the remaining 17~studies served as 
the HCD\@. This approach results in 17~HC data sets in which the average tumor incidence ($\hat{\pi}$) ranged 
between~0.0202 and~0.0270. The intra-class correlation ($\hat{\rho}$) was found to lie between~0.0179 
and~0.0288. Note that, due to the low tumor incidence, each study was comprised of at least one
group in which none of the rats developed a carcinoma. Therefore, all approaches that rely on the simple
GLM are prone to the Hauck-Donner effect and lead to similar results, regardless if HCD is considered or not.
Therefore, only the results for the simple GLM are considered below, but both pooling approaches are only shown
depending on a BGLM\@.

In the following steps, the data will first be analysed as they are, and second, CCG data will be changed artificially 
to illustrate effects of reducing sample size on real data.
If the studies' CCG remain unchanged, only in the first treatment group of study~3, the lower limits 
of all nine methods were above one, suggesting a significant increase over the control (numbers in 
Figure~\ref{fig::BfR_F344}). Further
significant increases of the tumor incidence were found in the second treatment group for both
frequentist pooling approaches, whereas in the third treatment group only the test-then-pool approach
yielded a lower limit above one. 

If the tumor incidence increases in at least one treatment group and the number of rats remains stable between the groups,
both frequentist borrowing approaches yield lower limits that are considerably higher than for all other methods. This increase 
is high enough to suggest that in studies~8 and~9 the risk-ratios exceed~1 for at least one treatment.

It is noteworthy that in study~8 both frequentist pooling approaches suggest a significant increase in the tumor rate of 
the first treatment group over the control. For naive pooling, the lower confidence limit is~1.07, whereas the corresponding
estimate for the risk ratio is~3.25. Similarly, the test-then-pool approach yields a lower limit of~1.27 and an estimate
for the risk ratio of~3.98.  However, the incidence in both groups is exactly the same (4~out of~50),
meaning that the concurrent data provide no evidence at all for increased incidence compared to control.

If no carcinoma occurs in a group and the number of rats is stable between the groups (as in studies~2, 13, 15 and~17)
all methods that depend on Bayesian borrowing yield the highest lower limits. Contrarily, lower limits
that are calculated based on a GLM ignoring HCD practically approach zero, which can be explained 
by the Hauck-Donner effect. Both frequentist borrowing methods as well as the Bayesian version of the GLM yield 
lower limits in between the limits that depend on Bayesian borrowing and the ones drawn from the simple GLM\@.

In studies~4, 10 and~18, the number of rats for which data is available for the intermediary dose groups is lower than for the control and high dose group.
In study~4, the Bayesian borrowing approaches yield limits that are comparable to the ones obtained with frequentist pooling, 
whereas in study~10, the Bayesian borrowing approaches yield the highest limits (around~0.6) for the risk-ratio between the third treatment 
group and the CCG\@.

\begin{figure}[H]
\centering
\includegraphics[width=1\columnwidth]{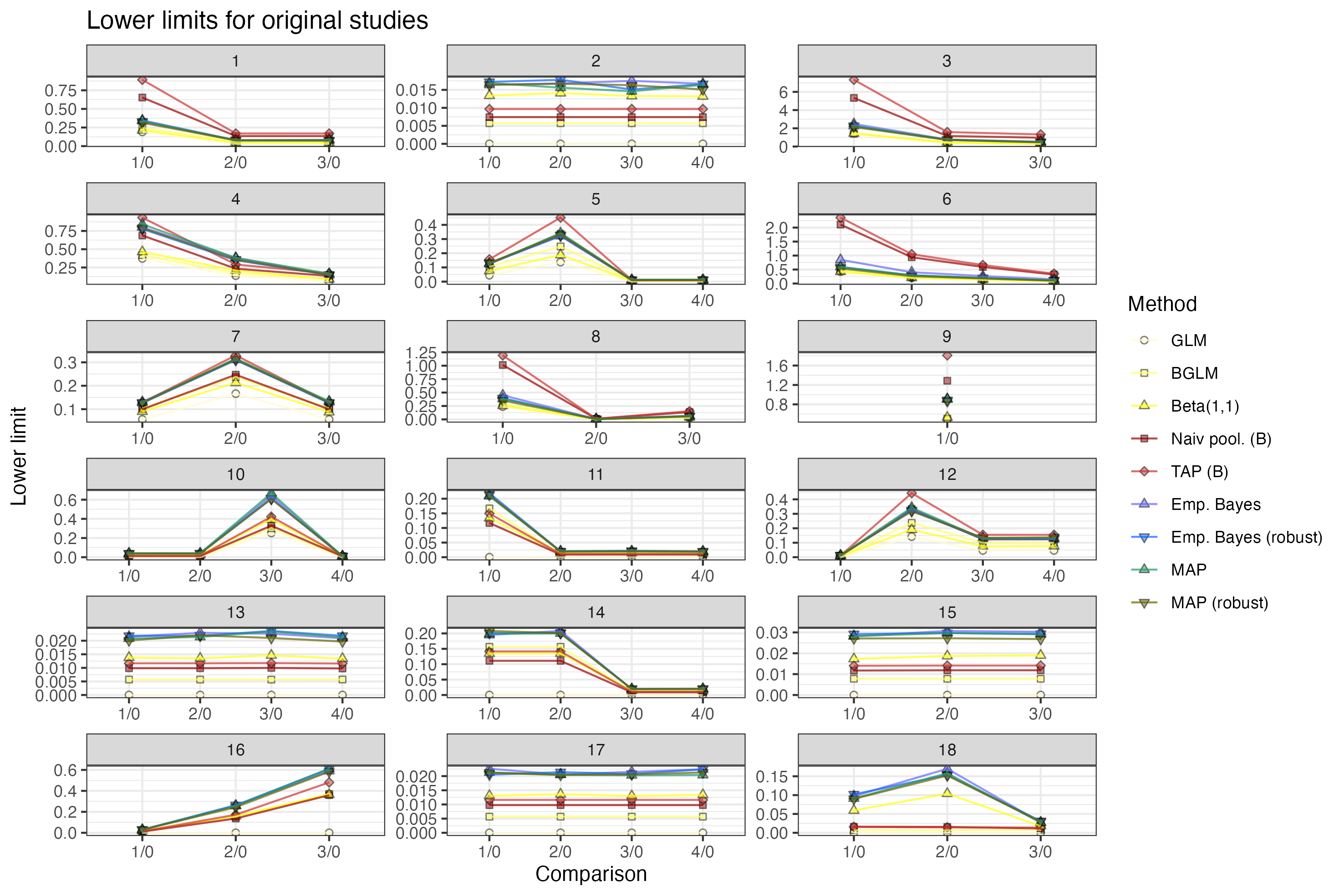}
\caption{Lower 95\% confidence or credible limits for risk ratios between treatment groups and the cocurrent control (studies as provided by the BfR).}
\label{fig::benchmark_original_studies}
\end{figure}

\begin{figure}[H]
\centering
\includegraphics[width=1\columnwidth]{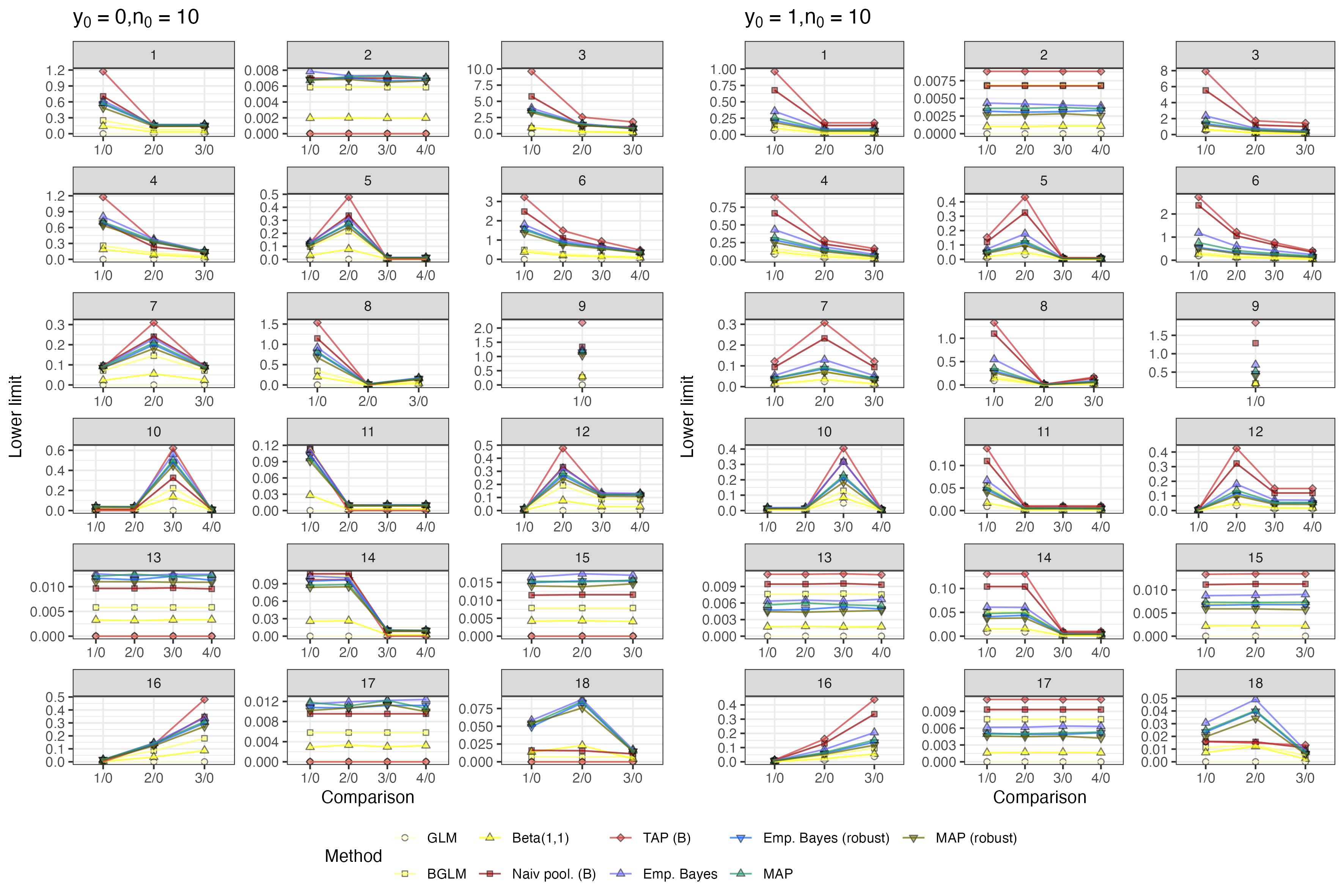}
\caption{Lower 95 \% limits if the CCG is changed (0/10, 1/10).
Black squares indicate lower limits obtained in the original data with the Bayesian GLM}
\label{fig::benchmark_lower_cl_10_01}
\end{figure}

In a second step, the data were reanalyzed with control groups artificially changed to~0 or~1 
rats with a carcinoma out of~10 rats per CCG, while the HCD remained with the original sample size. This was done to assess the
methods' behavior in the context of possible sample size reduction in the CCG. Since the tumor incidences in 
the original control groups ranged between~0 and~0.1, only 0/10 or 1/10 rats with a tumor in the reduced CCG 
reflect the properties of the original data (Figure~\ref{fig::benchmark_lower_cl_10_01}).

Though the general pattern remains stable, the number of lower limits that are found to be above~1 is increased
from~19 in the setting with unchanged CCG towards~32 in the setting in which the incidence is~0/10. Due to the small
sample size, none of the non-borrowing methods result in limits above~1. However, in the third study, all borrowing 
methods suggest significant increases in the tumor incidences of the first and the second treatment group. Also, for the 
first treatment group of study~6 and study~9, all borrowing methods result in lower limits above~1. 
In the first treatment group of studies~4 and~8 the frequentist methods result in limits above~1.
Compared to the analysis of the original data, the Bayesian approaches result in two additional studies 
that will be classified as positive, whereas frequentist pooling results in five further positive studies (compared to the unchanged setting).

If the incidence in the CCG is set to 1/10, all borrowing methods yield limits above~1 in the first treatment group
of study~3, whereas both frequentist borrowing approaches yield limits above~1 also for the second and third
treatment group. Together with the non-robustified empirical Bayes approach, both frequentist borrowing methods
yield limits above one in the first treatment group of study~6. In studies~8 and~9, both frequentist pooling 
approaches result in limits above one in the first treatment group. This means that in this setting, the empirical
Bayes approach results in one additional study that is classified as positive (compared to the unchanged setting),
whereas both frequentist pooling results in three additional positive studies.

%


\subsection{Application to EFSA example data}

\begin{figure}[H]
\centering
\includegraphics[width=1\columnwidth]{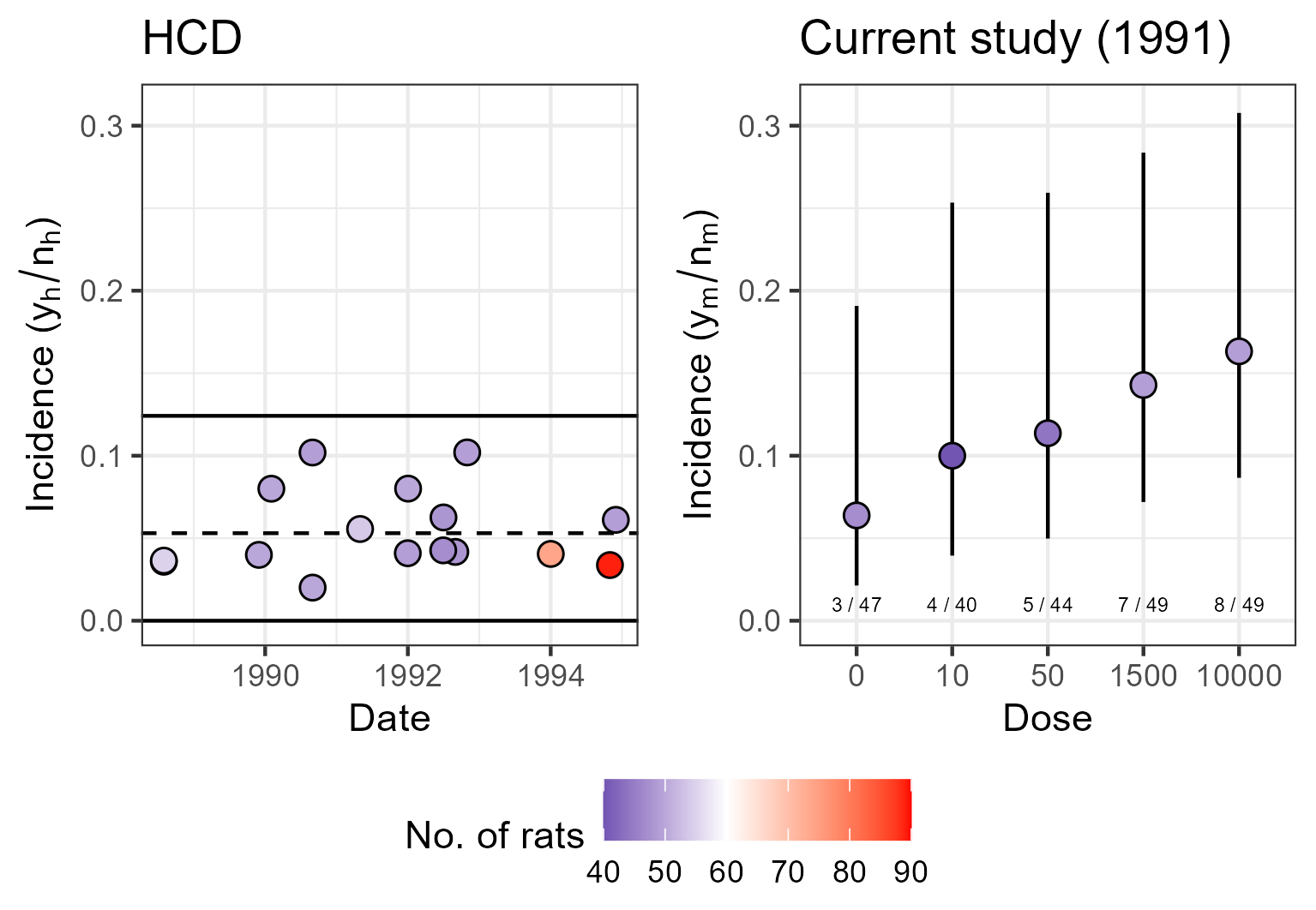}
\caption{Overview about the EFSA example data. 
Left panel: sample sizes~$n_h$ range between~47 and~89 in the HCD\@.
Black horizontal lines: 95\% prediction interval $[0,  0.125]$ for the observed success probability in the CCG\@.
Dashed horizontal line: Historical average success probability $\hat{\pi}=0.053$.
Numbers in right panel: Findings $y_m$ and sample size $n_m$.
Black vertical bars: Pointwise 95\% confidence intervals for the true success probabilities in the current trial.
}
\label{fig::EFSA_data}
\end{figure}

In their recently published scientific opinion on the use and reporting of historical control 
data for regulatory studies, the EFSA emphasized the inclusion of HCD in the test procedure
of the current trial (EFSA 2025). For binomial endpoints (e.g. tumor incidences), the EFSA used an example
on pathological findings (Tables~2 and~3 in annex~A of the opinion). To be in line with the analysis 
of the EFSA, all studies with a duration less than 104~weeks were not considered for analysis. 
The success probabilities for the remaining 16 historical studies, together with the ones 
obtained in a current trial are shown in Figure~\ref{fig::EFSA_data}. 
\newpage

Note that the estimate for the intra-class correlation 
is slightly negative $\hat{\rho}=-0.007$ indicating empirical underdispersion. However, 
a negative ICC implies a negative correlation between the individuals within each control group.
Practically spoken, this would mean that with the first occurrence of an animal with a finding,
the chance to obtain another test animal with the finding of interest decreases, which is biologically 
implausible. Furthermore, the estimate for the ICC is known to be negatively biased (Menssen and Rathjens 2025)
meaning that observing a negative estimate does not imply that the underlying process is truly underdispersed.
For this two reasons, the ICC was set to~$\hat{\rho}=0.00001$ in the subsequent analysis, such that the 
beta-binomial model (which does not allow for underdispersion) was applicable.

As an attempt to check for drift between HCD and CCG, it is required by the EFSA to evaluate
if the observed success probability of the current control ($y_0 / n_0$) falls within the 
95\% prediction interval obtained from the HCD\@. 
The prediction interval
for beta-binomial data applied here ($[0,  0.125]$, black horizontal lines in Figure~\ref{fig::EFSA_data}) was calculated
following Menssen and Rathjens 2025 using the \texttt{beta\_bin\_pi()} function from the R package \texttt{predint}. 
The observed success probability in the current control $\hat{\pi}_0=0.0638$ is 
relatively close to its historical average $\hat{\pi}=0.053$ and between-study variation is practically absent.

\begin{figure}[H]
\centering
\includegraphics[width=1\columnwidth]{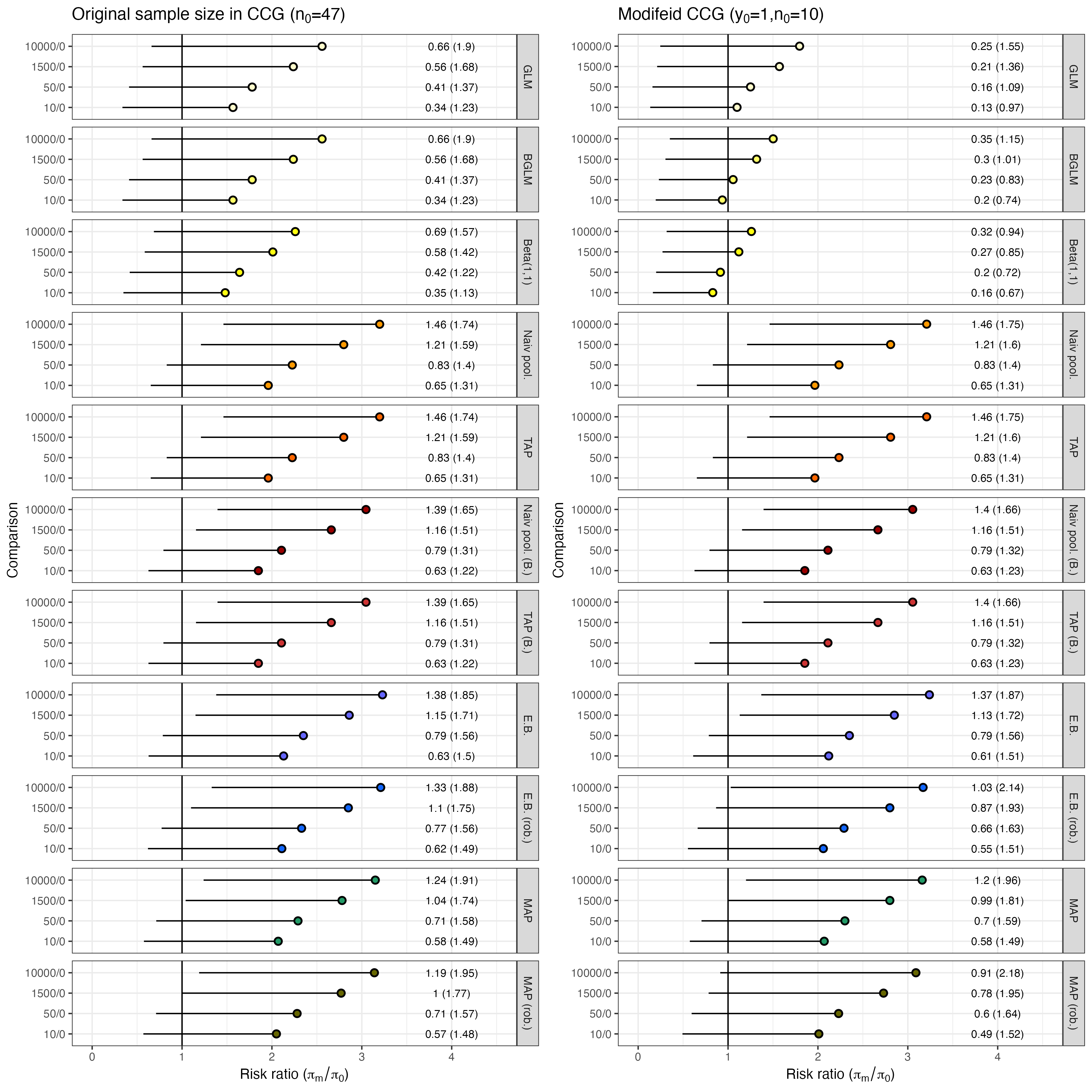}
\caption{Estimates for risk ratios and their corresponding simultaneous lower 95\% confidence or credible limits.  
Numbers: Lower limits.
Numbers in brackets: Distance from lower limit to the estimated ratio.}
\label{fig::EFSA_ci}
\end{figure}

In a first step, all methods were applied to the original data and simultaneous confidence or credible lower limits
for the risk ratios~$\pi_m / \pi_0$ ($m > 0$) were derived as described in the sections above. 
The corresponding estimates and lower limits are given in the left panel of Figure~\ref{fig::EFSA_ci}. 
If one does not consider HCD, no significant increase of tumor incidence in dose groups compared to control is found,
because the null hypotheses $H_{0m} = \pi_m / \pi_0 =1$ ($ m > 0$) is always covered by the lower bounds. 

\newpage

At first glance, all intervals that depend on borrowing look relatively similar 
and suggest an increased success probability for the two highest dose groups over the control. 
Both frequentist methods that rely on simple GLM result in exactly the same confidence limits. The same is true for borrowing based on bayesian GLM. This is because the success probability
in the CCG does not significantly differ from the incidences found in the historical control groups, regardless of the type of GLM (Bayesian or frquentist). 
Furthermore,
pooling using frequentist GLM yields the highest lower limits and the shortest intervals (here defined as the distance between the point estimates for the risk ratios and the lower limits).  If 
pooling is performed based on a Bayesian GLM, the lower limits are only slightly higher than the ones computed based on the empirical Bayes approach.

\begin{figure}[H]
\centering
\includegraphics[width=1\columnwidth]{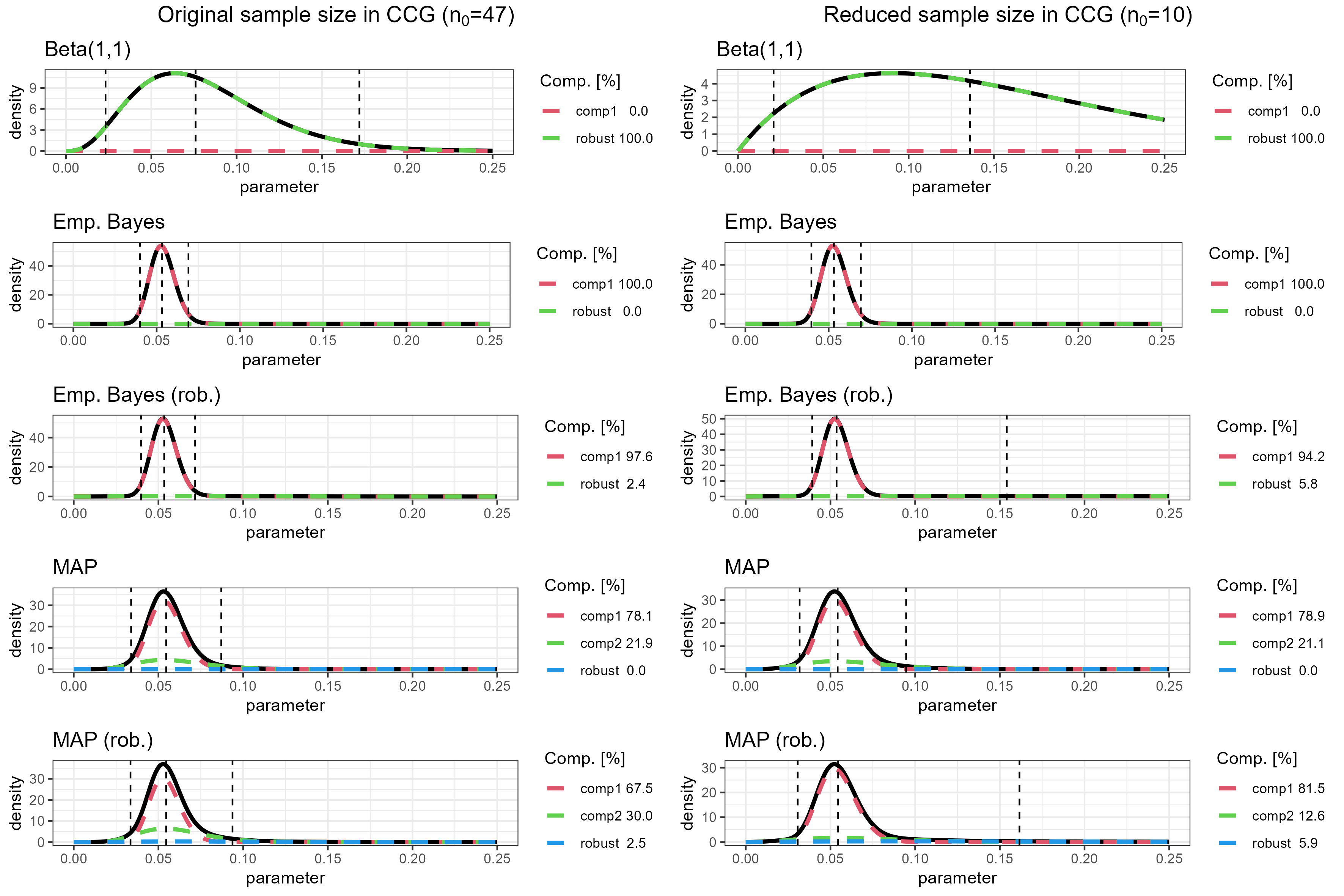}
\caption{Posterior distributions for the success probabilities $\pi_0$ in the CCG with original sample 
size (left column) and reduced sample size (right column) based on different prior distributions.
Solid black curves: Posterior distributions for $\pi_0$.
Dashed colored curves: Components of the posterior distributions.
Dashed vertical lines: 2.5\%, 50\% and 97.5\% quantiles of the posterior. 
}
\label{fig::EFSA_post}
\end{figure}

The MAP approach yields slightly lower limits and shorter intervals (at least for the comparisons for the two higher dose groups) 
than the empirical Bayes procedure. Robustification slightly lowers the limits and increases interval width. However, due to
the fact that systematic between-study variability (overdispersion) is absent in the HCD and the success probability in the CCG
is close to the historical average incidence, the uninformative component of the robustified priors is heavily down-weighted, 
($\tilde{\omega}^{rob EB}=0.024$ and $\tilde{\omega}^{rob MAP}=0.025$; see Figure~\ref{fig::EFSA_post}). Due to this heavy down-weighting of the uninformative component, the posterior distribution almost entirely resembles the informative prior component.

In order to demonstrate the possibility of augmentation and its effects on inference, the analysis was also done with a ``current'' 
data set in which the success probability of the control group was artificially changed to $y_0/n_0 = 1 / 10 = 0.1$ (right panel 
of Figure~\ref{fig::EFSA_ci}). Though at first glance, the overall picture does not change dramatically, the reduced sample size in
the CCG and the slightly higher point estimate for its success probability result in credible limits that are slightly lower
compared to the original setting. This can be explained by an increase of uncertainty about the success probability in the CCG,
which results in an increase of posterior variance (see Figure~\ref{fig::EFSA_post}) and hence, in a loss of precision. However, except for the robustified MAP-approach, an increase of the success probability in the highest dose group over the control can still be claimed.

\subsubsection{Reporting of HCD use}

It has to be emphasized that the Bayesian borrowing approaches discussed here, are transparent tools 
for the application of HCD in the context of a current study. One can report the parameters for the 
prior distribution (Table~\ref{tab::MAP_prior_par}) and its characteristics (e.g. its effective sample size of~227
and its mean of~0.056). Once the study is run, the posterior distribution can be reported similarly (Table~\ref{tab::MAP_posterior}).
For the example in which the CCG had 1 / 10 findings, the posterior median becomes~0.055 and the posterior effective sample size is~285.
Note that the high increase of the posterior effective sample size can be explained by the fact that the uninformative
component is heavily down-weighted in the posterior ($\tilde{\omega}^{rob}=0.059$).

\begin{table}[H]
	\centering
	\caption{Parameters of the robustified MAP prior} 
\begin{tabular}{lccc}
	\hline 
	 & \textbf{Weight} & \textbf{a} & \textbf{b} \\ 
	\hline 
	Component 1 &  0.6494 & 23.1972 & 411.6442 \\ 
	Component 2 & 0.1506 & 3.9494 & 61.46489 \\ 
	$\betadistn(1, 1)$ & 0.2 & 1.0 & 1.0 \\ 
	\hline 
\end{tabular} 
\label{tab::MAP_prior_par}
\end{table}

\begin{table}[H]
	\centering
	\caption{Parameters of the posterior distribution for the tumor incidence in the modified CCG ($y_0=1$ and $n_0=10$).} 
\begin{tabular}{lccc}
	\hline 
	 & \textbf{Weight} & \textbf{a} & \textbf{b} \\ 
	\hline 
	Component 1 &  0.7688 & 24.1972 & 421.6442 \\ 
	Component 2 & 0.1720 & 4.9494 & 71.464 \\ 
	$\betadistn(1, 1)$ & 0.0591 & 2.0 & 11.0 \\ 
	\hline 
\end{tabular} 
\label{tab::MAP_posterior}
\end{table}


\subsubsection{Planning of HCD use}

It is possible to discuss the role of the prior distribution already in the
planing stage of a new study, e.g., to describe worst-case scenarios
and possible effects of borrowing in the case of drift.
For this it is important to recall
that robustification is ``dynamic'' in two stages of the borrowing process: First,
the variance of the prior is proportional to the between-study 
variance. This means that the prior effective sample size decreases with increasing 
between-study variance.
Second, robustification reflects the degree of skepticism about a potential
prior-data conflict (also called drift), meaning that weighting of the uninformative 
component in the robustified prior with 0.2 corresponds to account for a 20\% chance 
of prior-data conflict.

Based on the non-robustified prior (Table~\ref{tab::MAP_prior_nonrob}) it is possible 
to derive procedures to deal with cases in which the number of finding in the CCG is high enough to 
increase the prior weight for the uninformative component to e.g. 0.5 (50\% chance of data prior conflict).

\begin{table}[H]
	\centering
	\caption{Parameters of the non-robustified MAP prior} 
\begin{tabular}{lccc}
	\hline 
	 & \textbf{Weight} & \textbf{a} & \textbf{b} \\ 
	\hline 
	Component 1 &  0.8118 & 23.1972 & 411.6442 \\ 
	Component 2 & 0.1882 & 3.9494 & 61.46489 \\ 
	\hline 
\end{tabular} 
\label{tab::MAP_prior_nonrob}
\end{table}

\begin{figure}[H]
\centering
\includegraphics[width=1\columnwidth]{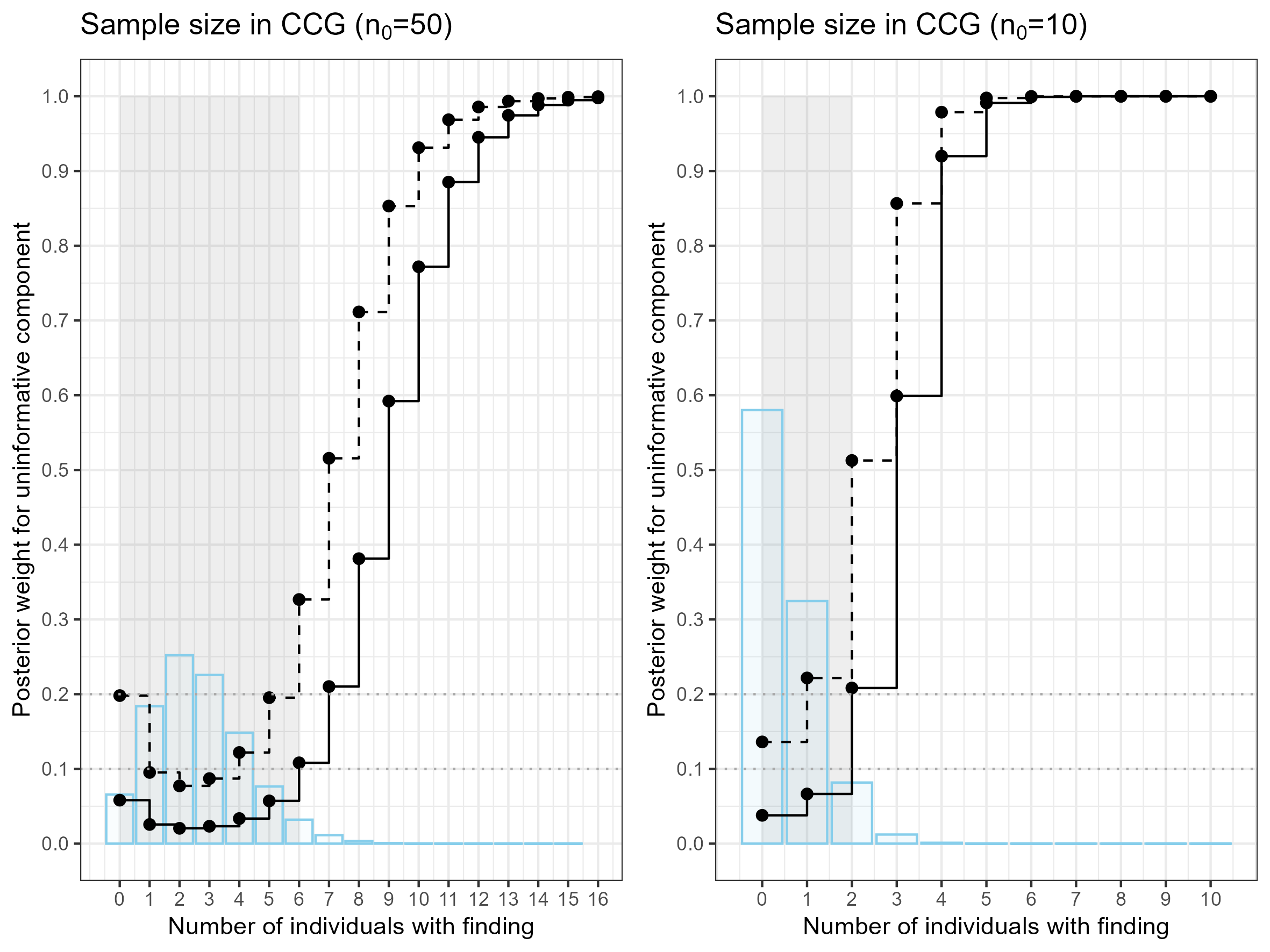}
\caption{Behaviour of the posterior weights for the uninformative component in different scenarios.
Solid line: Posterior weights if the prior weight for the uninformative component is set to 0.2.
Dashed line: Posterior weights if the prior weight for the uninformative component is set to 0.5.
Blue bars: prior predictive distribution.
Grey area: Central 95\% of the prior predictive distribution.}
\label{fig::EFSA_post_weights}
\end{figure}

Figure~\ref{fig::EFSA_post_weights} illustrates how the posterior weight for the uninformative component
changes depending on different numbers of possible findings in the future control group
and different prior weights $\omega^{rob}=0.2$ (solid line) and $\omega^{rob}=0.5$ (dashed line). 
The blue bars represent the prior predictive distribution which is a mixture of the 
non-robustified MAP prior and a binomial distribution. The grey area indicates its
central 95\%. 

If the sample size of the future control group is planned to be $n_0=50$, most historical information will be borrowed 
if the numbers of findings range between one and four. For 6~findings, the prior information is only slightly up-weighted,
whereas, down-weighting of the HCD (in relation to the prior weight) begins if 7~findings are present in the CCG\@. 
Comparing the predicted numbers of individuals with the weights attributed to informative and uninformative prior components, one can see that the implemented robustification seems to serve the purpose of downweighting prior information in case of an apparent prior-data conflict. 

\newpage

\section{Discussion}

The inclusion of HCD in the inference drawn from a current study is a highly discussed 
topic in the field of non- and pre-clinical toxicology and is also discussed by regulatory 
agencies such as the EFSA (EFSA 2025) and other stakeholders from industry, academia or non-governmental 
organizations (Golden et al. 2024). 
While in pre-clinical contexts, data from a range of similar, comparable experiments is often available, there are ethical responsibilities as well as economical incentives to fully utilize any available data and potentially reduce the amount of control data necessary.

As demonstrated above, information borrowing can be applied either to increase the power of the test 
or for augmentation of CCG animals while maintaining the power. However, it seems that for reduction of the number of used animals 
in pre-clinical toxicology, the focus is	laid on 	the direct replacement of CCG animals with historical ones matched 
as close as possible to the conditions of the current study, following the concept of virtual control groups
(Steger Hartmann et al. 2020, Gurjanov et al. 2024, Golden et al. 2024). 
	
	
Therefore, one of the key-findings of our study is 
that na\"ive borrowing
which is based on complete pooling 
of HCD and CCG animals leads to the highest FWER inflation (except for very small success probabilities 
$\pi=0.01$), even if between-study variability and / or drift is absent, meaning that HCD and CCG are
perfectly exchangable and relate to a single, \emph{common} underlying parameter.
This can be explained by the fact that for such kind pooling, the HCD are directly treated as additional 
observations in the current control without any 
allowance for potential \emph{heterogeneity} (among HCD or between HCD and CCG).
This means that the standard error of the pooled estimate $se(\hat{\pi}_0)$ decreases
with increasing sample size of the pooled control $n_0^{pool}$.

\emph{Hierarchical} or \emph{random-effects} models on the other hand allow for potential variability due to differences between HCD samples or between HCD and CCG data.
Consequently, a simple common-effect approach will likely yield inferences of over-optimistic precision, while hierarchical approaches may capture additional variation and yield more cautious predictions. In addition, \emph{robustification} approaches allow to account for additional discrepancies that may be anticipated between HCD and CCG data.
Similar results were obtained by Wang et al. 2022 who studied the augmentation of patients in randomized 
clinical trials by historical ones that were matched with patients from the current study via propensity scores,
an approach which is fairly close to the VCG concept of Steger-Hartmann et al. 2020.
Contrary, we could show, that in the absence of a data-prior conflict or rare events 
($\delta=1$ and $\pi \geq 0.1$), the Bayesian borrowing approaches that depend on robustified
prior distributions satisfactorily control the FWER\@. 

It was possible to show that 
even if the CCG sample size is reduced from~50 to~10, the approaches which depend on robustified priors were 
able to retain the power of the non-borrowing approaches that depend on the original CCG sample size.
However, it has to be noted that in their current setup, long-term carcinogenicity studies may remain heavily underpowered
and substantial gains in the power to detect at least one treatment group with increased toxicity 
(approx. 30 percentage points) are only possible in scenarios of very small between-study heterogeneity and very high 
increases in the success probability (see left hand side of Figure~\ref{fig::power_app_175}, $n_0=10$, $\rho=1e-05$). 
For smaller increases of success rates, the any-pairs power (APP) was
only moderately increased for all Bayesian borrowing approaches. These findings are in line with the power simulations
of Kitsche et al. 2012, who provided an asymptotic empirical Bayes approach to conduct a Dunnett test for 
differences in success probabilities.
	
However, the potential for animal reduction in the CCG and / or power gains through employment of more complex models and inclusion of additional data comes at the risk that any of the additional assumptions being made may also be violated. One example is the potential 
FWER inflation in the case of drift ($\delta > 1$ in Figure~\ref{fig::FWER}) meaning that the FWER can inflate
towards 100\% if pooling approaches or non-robustified priors are applied and the amount of drift is large enough.
The application of mixture priors that are robustified with a non-informative component (here 20\%), decrease 
the potential for FWER inflation to a maximum of 0.3, which is still substantial, but only reached if the sample size 
of the CCG is decreased ($n_0=10$) and the success probability is relatively high ($\pi=0.5$).  
At this point it is important to recall, that pre- and non-clinical toxicological studies are run in order to
determine if the compound of interest affects human health in a potentially harmful direction. From this point
of view, slight increases of the FWER might be acceptable if borrowing at the same time results in increased power.

We applied a number of proposed approaches to data from 18~real-life rat carcinogenecity studies examining a C-cell carcinoma --- a rare tumor type --- in a benchmarking exercise and found a similar picture. 
If the CCG remained unchanged, the frequentist pooling approaches resulted in three
additional positive studies (with at least one significant increase in at least one treatment group), of which
one was an obvious false-positive. Contrary, the Bayesian borrowing approaches came to the similar conclusion as
the non-borrowing approaches (only study~3 was positive).

However, if the CCG is artificially reduced in order to simulate a animal-saving reduction of the control group size, 
none of the non-borrowing approaches resulted in a positive study.
Furthermore, the conclusions regarding borrowing depend heavily on the incidences. 
If the incidence is low (0/10), all borrowing methods lead to the conclusion that three studies (studies~3, 6 and~9)
are positive. Additionally both frequentist pooling approaches result in two more positive studies (studies~4 and~8).
With increasing incidence, the number of studies that are found to be positive with Bayesian approaches decreased dramatically:
For 1/10 only study three remained positive (for all Bayesian approaches) and study~6 was found to be positive for the 
non-robustified empirical Bayes approach. On the other hand, studies~3, 6, 8 and~9 remained positive if frequentist
pooling was applied.

Though the benchmark analysis provided first insights into real-life application, it has to be pointed out 
that in a typical long-term carcinogenicity study approx. 40--50 organs and tissues, all able to generate 
different types of tumors (in total around 300--500 distinct neoplastic changes) are assessed 
simultaneously and study conclusions are not solely based on statistical significance (OECD 2018). Since our analysis was
run based on one single endpoint, these important points are not considered and remain as an open issue
for future research.

To our knowledge, the literature on testing procedures that account for HCD and adjust
for multiplicity is extremely limited and the application of robustified mixture priors to multiple testing
procedures has not been demonstrated before. Note that the rank based method of Besag et al. 1995 which was
applied here for adjustment is relatively flexible with regard to the applied distributions and inferential
problems, as long as the posterior is continuous. Therefore this study can be seen as a first attempt towards
multiple testing procedures that account for HCD\@. 
	
Next possible steps will be the application of mixture-priors in the context of trend-tests that are usually
part of the toxicological test battery (Dertinger et al. 2023) and other scales (e.g. count data or continuous endpoints)
as well as their adaption to account for several levels of hierarchy within a control group (e.g. to account for the fact that 
a certain group of animals is housed within a certain cage). 
	 
Furthermore, it has to be noted that the proposed Bayesian borrowing approaches are evaluated 
according to their frequentist operation characteristics only. We are aware, that this is somewhat contradictory 
from a strictly Bayesian point of view. Note that this is a general issue in the context of Bayesian borrowing
approaches. In order to overcome this problem, Best et al. 2025 proposed a Bayesian metric for two-sample
designs of clinical trials (average Type-1 error). However, this approach needs further extensions to be 
applicable in multiple testing scenarios as well.

\section{Acknowledgements}
We have to thank Sören Budig (Department of Biostatistics, Leibniz University Hannover) for the excellent technical support. 


\section{References}

Besag, J., Green, P., Higdon, D., \& Mengersen, K. (1995). Bayesian computation and stochastic systems. 
Statistical science, 3-41.\\

Best, N., Ajimi, M., Neuenschwander, B., Saint-Hilary, G., \& Wandel, S. (2025). Beyond the classical type I error: 
Bayesian metrics for Bayesian designs using informative priors. Statistics in biopharmaceutical research, 17(2), 183-196. \\

Chen, R., Wei, L., Dai, Y., Wang, Z., Yang, D., Jin, M., ... \& Zhong, N. (2024). Efficacy and safety of mepolizumab in a 
Chinese population with severe asthma: a phase III, randomised, double-blind, placebo-controlled trial. ERJ open research, 
10(3). \\

Dertinger, S. D., Li, D., Beevers, C., Douglas, G. R., Heflich, R. H., Lovell, D. P., ... \& Zhou, C. (2023). Assessing
the quality and making appropriate use of historical negative control data: A report of the International Workshop on 
Genotoxicity Testing (IWGT). Environmental and molecular mutagenesis, 1–22. \\

Djira G, Hasler M, Gerhard D, Segbehoe L, Schaarschmidt F (2025):  mratios: Ratios of Coefficients in the General Linear Model.
doi:10.32614/CRAN.package.mratios \\

EFSA Scientific Committee, Hardy, A., Benford, D., Halldorsson, T., Jeger, M. J., Knutsen, H. K., ... \& Alexander, J. (2017).
Guidance on the assessment of the biological relevance of data in scientific assessments. Efsa Journal, 15(8), e04970. \\

EFSA Panel on Plant Protection Products and their Residues (PPR), Coja, T., Adriaanse, P., Choi, J., Finizio, A., Giraudo, M., 
... \& Wilks, M. (2025). Use and reporting of historical control data for regulatory studies. EFSA Journal, 23(8), e9576. \\

Gart, J. J., Chu, K. C., \& Tarone, R. E. (1979). Statistical issues in interpretation of chronic bioassay tests for carcinogenicity. Journal of the National Cancer Institute, 62(4), 957-974.\\

Gelman A, Su Y (2024). arm: Data Analysis Using Regression and Multilevel/Hierarchical Models. doi:10.32614/CRAN.package.arm \\

Golden, E., Allen, D., Amberg, A., Anger, L. T., Baker, E., Baran, S. W., ... \& Steger-Hartmann, T. (2024). Toward implementing
virtual control groups in nonclinical safety studies: Workshop report and roadmap to implementation. ALTEX-Alternatives to 
animal experimentation, 41(2), 282-301.\\

Greim, H., Gelbke, H. P., Reuter, U., Thielmann, H. W., \& Edler, L. (2003). Evaluation of historical control data in
carcinogenicity studies. Human \& experimental toxicology, 22(10), 541-549. \\

Gurjanov, A., Vaas, L. A., \& Steger-Hartmann, T. (2024). The road to virtual control groups and the importance of proper 
body-weight selection. ALTEX 41(4):660-665.\\

Gurjanov, A., Kreuchwig, A., Steger-Hartmann, T., \& Vaas, L. A. I. (2023). Hurdles and signposts on the road to virtual 
control groups—a case study illustrating the influence of anesthesia protocols on electrolyte levels in rats. Frontiers 
in Pharmacology, 14, 1142534.\\

Hobbs, B. P., Carlin, B. P., Mandrekar, S. J., \& Sargent, D. J. (2011). Hierarchical commensurate and power prior models for adaptive incorporation of historical information in clinical trials. Biometrics, 67(3), 1047-1056.\\

Hothorn, T., Bretz, F., \& Westfall, P. (2008). Simultaneous inference in general parametric models. Biometrical Journal: 
Journal of Mathematical Methods in Biosciences, 50(3), 346-363. \\

Ibrahim, J. G., \& Chen, M. H. (2000). Power prior distributions for regression models. Statistical Science, 46-60.\\

Kitsche, A., Hothorn, L. A., \& Schaarschmidt, F. (2012). The use of historical controls in estimating simultaneous 
confidence intervals for comparisons against a concurrent control. Computational Statistics \& Data Analysis, 56(12), 
3865-3875.\\

Kluxen, F. M., Weber, K., Strupp, C., Jensen, S. M., Hothorn, L. A., Garcin, J. C., \& Hofmann, T. (2021). 
Using historical control data in bioassays for regulatory toxicology. Regulatory Toxicology and Pharmacology, 125, 105024. \\

Lenth R, Piaskowski J (2025). emmeans: Estimated Marginal Means, aka Least-Squares Means. doi:10.32614/CRAN.package.emmeans \\
  
Lui, K. J., Mayer, J. A., \& Eckhardt, L. (2000). Confidence intervals for the risk ratio under cluster sampling based on the 
beta‐binomial model. Statistics in medicine, 19(21), 2933-2942.\\

Mandel, M., \& Betensky, R. A. (2008). Simultaneous confidence intervals based on the percentile bootstrap approach. 
Computational statistics \& data analysis, 52(4), 2158-2165. \\

Menssen, M. (2023). The calculation of historical control limits in toxicology: Do's, don'ts and open issues from 
a statistical perspective. Mutation Research/Genetic Toxicology and Environmental Mutagenesis, 892, 503695. \\

Menssen M (2025): predint: Prediction Intervals. doi:10.32614/CRAN.package.predint \\

Menssen, M., \& Schaarschmidt, F. (2019). Prediction intervals for overdispersed binomial data with application to 
historical controls. Statistics in Medicine, 38(14), 2652-2663. \\

Menssen, M., \& Schaarschmidt, F. (2022). Prediction intervals for all of M future observations based on linear random effects models. Statistica Neerlandica, 76(3), 283-308.\\

Menssen, M., Dammann, M., Fneish, F., Ellenberger, D., \& Schaarschmidt, F. (2025). Prediction Intervals for Overdispersed 
Poisson Data and Their Application in Medical and Pre‐Clinical Quality Control. Pharmaceutical Statistics, 24(2), e2447.\\

Menssen, M., \& Rathjens, J. (2025). Prediction intervals for overdispersed binomial endpoints and their application 
to toxicological historical control data. Pharmaceutical Statistics, 24(5), e70033.\\

Neuenschwander, B., Weber, S., Schmidli, H., \& O'Hagan, A. (2020). Predictively consistent prior effective sample sizes. Biometrics, 76(2), 578-587.\\

OECD (2018), Test No. 451: Carcinogenicity Studies, OECD Guidelines for the Testing of Chemicals, Section 4, OECD Publishing, Paris. \\

R Core Team (2025). R: A Language and Environment for Statistical Computing. R Foundation for Statistical Computing, Vienna,
Austria. \\

Richeldi, L., Azuma, A., Cottin, V., Hesslinger, C., Stowasser, S., Valenzuela, C., ... \& Maher, T. M. (2022). Trial of a
preferential phosphodiesterase 4B inhibitor for idiopathic pulmonary fibrosis. New England Journal of Medicine, 
386(23), 2178-2187. \\

Röver, C., Wandel, S., \& Friede, T. (2019). Model averaging for robust extrapolation in evidence synthesis. 
Statistics in medicine, 38(4), 674-694.\\

Sanz, F., Pognan, F., Steger-Hartmann, T., Díaz, C., Asakura, S., Amberg, A., ... \& Wilkinson, D. (2023). 
eTRANSAFE: data science to empower translational safety assessment. Nature Reviews Drug Discovery, 22(8), 605-606.\\

SATO, G., NAKAJIMA, M., SAKAI, K., TOGASHI, Y., YAMAMOTO, M., INOUE, Y., ... \& SUZUKI, M. (2024). Potential issues associated 
with the introduction of virtual control groups into non-clinical toxicology studies. Translational and Regulatory Sciences, 
6(1), 1-9.\\

Schaarschmidt F (2018): BSagri: Safety Assessment in Agricultural Field Trials. doi:10.32614/CRAN.package.BSagri\\ 

Secrest M, Gravestock I (2025): psborrow2: Bayesian Dynamic Borrowing Analysis and Simulation.
doi:10.32614/CRAN.package.psborrow2 \\
  
Sato, G., Kurooka, T., Takakura, I., Amano, Y., Sakai, K., Nishikawa, S., ... \& Suzuki, M. (2025). Interfacility Variations in Blood Chemistry Parameters and Their Implications for Virtual Control Group Construction. NAM Journal, 100073.\\

Schmidli, H., Gsteiger, S., Roychoudhury, S., O'Hagan, A., Spiegelhalter, D., \& Neuenschwander, B. (2014). 
Robust meta-analytic-predictive priors in clinical trials with historical control information. Biometrics, 70(4), 1023-1032. \\

Smythe, R. T., Krewski, D., \& Murdoch, D. (1986). The use of historical control information in modelling dose response
 relationships in carcinogenesis. Statistics \& probability letters, 4(2), 87-93. \\

Steger-Hartmann, T., Kreuchwig, A., Vaas, L., Wichard, J., Bringezu, F., Amberg, A., ... \& Barber, C. (2020). 
Introducing the concept of virtual control groups into preclinical toxicology animal testing. Altex, 37(3), 343-349.\\

Steger-Hartmann, T., Sanz, F., Bringezu, F., \& Soininen, I. (2025). IHI VICT3R: developing and implementing virtual 
control groups to reduce animal use in toxicology research. Toxicologic Pathology, 53(2), 230-233.\\

Tarone, R. E. (1982a). The use of historical control information in testing for a trend in proportions. Biometrics, 
38(1), 215–220. \\

Tarone, R. E. (1982b). The use of historical control information in testing for a trend in Poisson means. Biometrics, 
38(2), 457–462. \\

Valverde-Garcia, P., Springer, T., Kramer, V., Foudoulakis, M., \& Wheeler, J. R. (2018). An avian reproduction study 
historical control database: A tool for data interpretation. Regulatory Toxicology and Pharmacology, 92, 295-302. \\

Viele, K., Berry, S., Neuenschwander, B., Amzal, B., Chen, F., Enas, N., ... \& Thompson, L. (2014). Use of historical 
control data for assessing treatment effects in clinical trials. Pharmaceutical statistics, 13(1), 41-54.\\

Wang, X., Suttner, L., Jemielita, T., \& Li, X. (2022). Propensity score-integrated Bayesian prior approaches for augmented 
control designs: a simulation study. Journal of Biopharmaceutical Statistics, 32(1), 170-190.\\

Wang, X., Dormont, F., Lorenzato, C., Latouche, A., Hernandez, R., \& Rouzier, R. (2023). Current perspectives for 
external control arms in oncology clinical trials: Analysis of EMA approvals 2016–2021. Journal of Cancer Policy, 35, 100403. \\

Weber, S., Li, Y., Seaman III, J. W., Kakizume, T., \& Schmidli, H. (2021). Applying meta-analytic-predictive priors with the R
Bayesian evidence synthesis tools. Journal of Statistical Software, 100, 1-32. \\

Weber, S. (2025). Getting started with RBesT (binary). \url{https://cran.r-project.org/web/packages/RBesT/vignettes/introduction.html}. Assessed 13.02.2026.\\

Zarn, J. A., König, S. L., Shaw, H. V., \& Geiser, H. C. (2024). An analysis of the use of historical control data in 
the assessment of regulatory pesticide toxicity studies. Regulatory Toxicology and Pharmacology, 154, 105724.\\

Zheng, H., \& Hampson, L. V. (2020). A Bayesian decision‐theoretic approach to incorporate preclinical information into phase I
oncology trials. Biometrical Journal, 62(6), 1408-1427.\\

\end{document}